\documentclass[aps,prd,showpacs,nobibnotes,superscriptaddress,nofootinbib,twocolumn]{revtex4-1}
\usepackage{epsfig,epstopdf}
\usepackage{amsmath,amssymb,amsfonts,multirow,graphicx}

\DeclareMathOperator{\tr}{Tr}

\begin{document}

\title{Minimal anomaly-free chiral fermion sets and gauge coupling unification}

\author{Lu\'{\i}s M. Cebola}
\email{luismcebola@tecnico.ulisboa.pt}
\affiliation{Departamento de F\'{\i}sica and Centro de F\'{\i}sica
  Te\'{o}rica de Part\'{\i}culas (CFTP), Instituto Superior T\'{e}cnico,
  Universidade de Lisboa, Av. Rovisco Pais, 1049-001 Lisboa,
  Portugal}

\author{D.~Emmanuel-Costa}
\email{david.costa@tecnico.ulisboa.pt}
\affiliation{Departamento de F\'{\i}sica and Centro de F\'{\i}sica
  Te\'{o}rica de Part\'{\i}culas (CFTP), Instituto Superior T\'{e}cnico,
  Universidade de Lisboa, Av. Rovisco Pais, 1049-001 Lisboa,
  Portugal}

\author{R.~Gonz\'{a}lez Felipe}
\email{ricardo.felipe@tecnico.ulisboa.pt}
\affiliation{Instituto Superior de Engenharia de Lisboa - ISEL,
 Rua Conselheiro Em\'{\i}dio Navarro 1, 1959-007 Lisboa,
 Portugal
}
\affiliation{Departamento de F\'{\i}sica and Centro de F\'{\i}sica
  Te\'{o}rica de Part\'{\i}culas (CFTP), Instituto Superior T\'{e}cnico,
  Universidade de Lisboa, Av. Rovisco Pais, 1049-001 Lisboa,
  Portugal}

\author{C.~Sim\~{o}es}
\email{catarinafsimoes@tecnico.ulisboa.pt}
\affiliation{Departamento de F\'{\i}sica and Centro de F\'{\i}sica
  Te\'{o}rica de Part\'{\i}culas (CFTP), Instituto Superior T\'{e}cnico,
  Universidade de Lisboa, Av. Rovisco Pais, 1049-001 Lisboa,
  Portugal}

\pacs{12.10.-g, 14.80.-j, 11.10.Hi}

\begin{abstract}
We look for minimal chiral sets of fermions beyond the standard model that are anomaly free and, simultaneously, vectorlike particles with respect to color $\mathsf{SU}(3)$ and electromagnetic $\mathsf{U}(1)$. We then study whether the addition of such particles to the standard model particle content allows for the unification of gauge couplings at a high energy scale, above $5.0\times10^{15}$ GeV so as to be safely consistent with proton decay bounds. The possibility to have unification at the string scale is also considered. Inspired in grand unified theories, we also search for minimal chiral fermion sets that belong to $\mathsf{SU}(5)$ multiplets, restricted to representations up to dimension~$50$. It is shown that, in various cases, it is possible to achieve gauge unification provided that some of the extra fermions decouple at relatively high intermediate scales.
\end{abstract}

\maketitle

\section{Introduction}
\label{sec:intro}

The standard model~(SM) of particle physics is a very successful theory that describes the fundamental known particles and their interactions based on the gauge symmetry $\mathsf{SU}(3) \times \mathsf{SU}(2) \times \mathsf{U}(1)$. So far, the discovered elementary fermions are chiral Weyl particles and thus cannot have any gauge-invariant mass term. Once the SM gauge symmetry is spontaneously broken at the electroweak scale, fermions acquire masses and the remaining symmetry gives rise to interactions whose nature is vectorlike with respect to color $\mathsf{SU}(3)$ and electromagnetic $\mathsf{U}(1)$.

In the SM, the strong $\alpha_s$, weak isospin $\alpha_w$ and hypercharge $\alpha_y$ gauge couplings are not related among themselves by any symmetry principle. It is well known that the gauge couplings $\alpha_{1,2,3}\equiv \kappa_{1,2,3}\,\alpha_{y,w,s}$, where $\kappa_i$ are normalization constants, evolve with the energy scale according to the renormalization group equations (RGEs). At one-loop level, one verifies that $\alpha_1$ and $\alpha_2$ unify around $10^{13}$ GeV, while $\alpha_2$ and $\alpha_3$ unify at $10^{17}$ GeV for the canonical normalization $\kappa_1=5/3$ and $\kappa_2=\kappa_3=1$, like in $\mathsf{SU}(5)$ or $\mathsf{SO}(10)$ group. This fact already suggests a possible unification of the three couplings, either by changing the normalization factor $\kappa_1$ associated to the hypercharge, or by adding new particle content that properly modifies the running of the couplings. In the former case, complete unification is achieved for $\kappa_1\approx13/10$, assuming that the SM is an effective theory valid up to the unification scale around $10^{17}$ GeV. It is interesting that this scale is close to the scale predicted in string theories~\cite{Dienes:1996du}. In the second case, it has been noted that extending the SM with a fourth generation of quarks and leptons gives freedom for the convergence of gauge couplings to a common value at a scale around $3 \times 10^{15}$~GeV~\cite{Hung:1997zj}.

Adding new chiral fermions to the SM particle content unavoidably raises the question of gauge anomaly cancellation. In general, this is not a trivial issue, since the anomaly-free conditions depend on the transformation properties of the new fermions under the SM gauge group. Furthermore, it is also necessary to guarantee that after electroweak symmetry breaking the theory remains vectorlike with respect to color $\mathsf{SU}(3)$ and electromagnetic $\mathsf{U}(1)$, in order to conserve parity. Remarkably, these properties are verified between quarks and leptons within each generation of the SM~\cite{Gross:1972pv, Bouchiat:1972iq, Georgi:1972bb}. On the other hand, adding only vectorlike fermions to the theory does not bring any anomaly constraint, since their contribution to the gauge anomalies adds to zero. Another motivation for introducing vectorlike particles is that it is possible to construct gauge-invariant mass terms for them, and the masses are not necessarily below the electroweak scale, implying that these particles can decouple from the low-energy spectrum of the theory. These are among the reasons for following this path for gauge coupling unification in several extensions of the SM (see, e.g., Refs.~\cite{Amaldi:1991zx,EmmanuelCosta:2005nh,EmmanuelCosta:2006rr,Gogoladze:2010in,palash,Perez:2014kfa}).

Concerning the possibility of extending the SM with anomaly-free chiral fermion sets, one can find numerical analyses~\cite{Eichten:1982pn} and theoretical studies~\cite{Fishbane:1983hf,Fishbane:1984zv,Foot:1988qx,Frampton:1993bp,Batra:2005rh} in the literature. In particular, patterns for adding anomaly-free charge-vectorial chiral sets of fermions that acquire mass through their coupling to the SM Higgs doublet have been investigated in Refs.~\cite{Fishbane:1983hf,Fishbane:1984zv}. Furthermore, a general structure of exotic generations and the corresponding analysis of possible quantum numbers were presented in Ref.~\cite{Foot:1988qx}, assuming the gauge and Higgs structure of the SM. Finally, a complete description of chiral anomaly-free sets for gauge theories with additional $\mathsf{U}(1)$ groups can be found in Ref.~\cite{Batra:2005rh}; however, the authors restricted their analysis to chiral sets that only include vectorlike fermions with respect to non-Abelian groups. We also remark that in the above works the problem of gauge coupling unification has not been addressed.

The aim of our work is to extend previous studies in order to find minimal chiral sets of fermions beyond the SM that are not only anomaly free but also lead to gauge coupling unification at high energies, inspired by grand unified theories (GUTs) or string theories. The paper is organized as follows. In Sec.~\ref{sec:arbitrary}, we study the problem of finding the minimal sets of chiral fermions beyond the SM that lead to anomaly-free solutions. We then proceed in Sec.~\ref{subsec:GCU} to explore the sets that lead to the unification of the gauge couplings at one-loop level. In our analysis, we restrict to cases in which the unification scale is above $5.0\times10^{15}$ GeV, so as to be safely consistent with proton decay bounds~\cite{Nath:2006ut}. Furthermore, we allow for the decoupling of the beyond-the-SM fermions at some intermediate energy scale between the electroweak and the unification scales. We also look for solutions in which gauge and gravitational couplings match at a unique (string) scale. We then proceed in Sec.~\ref{sec:SU5}, inspired by GUTs, to study the possibility of having anomaly-free solutions with chiral fermion multiplets belonging to $\mathsf{SU}(5)$ representations up to dimension 50. For the anomaly-free sets obtained, we  analyze whether the gauge couplings unify at GUT and/or string energy scales. Finally, our conclusions are presented in Sec.~\ref{sec:conclusions}.

\section{Minimal anomaly-free chiral fermion sets}
\label{sec:arbitrary}

In this work, we search for new chiral sets of particles beyond the SM that are anomaly free and at the same time QED and QCD vectorlike at low energies (i.e., electric and color currents are vectorlike), so that parity symmetry is not spoiled. Without loss of generality, we assume that all chiral fields are left handed. To be free of anomalies, the new chiral fermions must verify the following anomaly conditions with respect to the SM gauge group:
\begin{subequations}
\label{eq:anomaly}
\begin{align}
\left[\mathsf{SU}(3) \,\mbox{-}\,\mathsf{SU}(3)\, \mbox{-}\,\mathsf{SU}(3)\right] &:\quad \sum_R A_3(R)\,d_2(R) \,=\, 0\,,\\
\left[\mathsf{SU}(3) \,\mbox{-}\,\mathsf{SU}(3)\, \mbox{-}\,\mathsf{U}(1)\right] &:\quad \sum_R y_R\,t_3(R)\,d_2(R) \,=\, 0\,,\\
\left[\mathsf{SU}(2) \,\mbox{-}\,\mathsf{SU}(2)\, \mbox{-}\,\mathsf{U}(1)\right] &:\quad \sum_R y_R\,t_2(R)\,d_3(R) \,=\, 0\,,\\
\left[\mathsf{U}(1) \,\mbox{-}\,\mathsf{U}(1)\, \mbox{-}\,\mathsf{U}(1)\right] &:\quad \sum_R y^3_R\,d_2(R)\,d_3(R) \,=\, 0\,,\\
\left[\text{gravity} \,\mbox{-}\,\text{gravity}\, \mbox{-}\,\mathsf{U}(1)\right] &:\quad \sum_R y_R\,d_2(R)\,d_3(R) \,=\, 0\,,
\end{align}
\end{subequations}
where $d_i(R)$, $A_i(R)$,  and  $t_i(R)$ are respectively the dimension, cubic anomaly index, and Dynkin index of the representation $R$ with respect to the subgroup $G_i$. For a given representation $R$ of a group $G$, the integer cubic index $A(R)$ is defined through the relation~\cite{Banks:1976yg}
\begin{equation}
\label{eq:Aconstant}
\tr\left(\left\{T^a_R,T^b_R\right\}T^c_R\right)\,= \,A(R)\,\tr\left(\left\{T^a,T^b\right\}T^c\right),
\end{equation}
where $T_R^a$ are the generators of $G$ for the representation $R$, and $T^a$ for the fundamental representation. Clearly, for the conjugate representation $\overline{R}$ one has $A(\overline{R})=-A(R)$, which in turn implies $A(R)=0$ for any real representation. The Dynkin index of a representation $R$ can be determined through the Dynkin labels~\cite{Fischler:1980ck,Slansky:1981yr} as
\begin{equation}
\label{eq:C1}
\begin{split}
t_m(R)=&\frac{d_m(R)}{2m(m^2-1)}\left[\sum_{j=1}^{m-1} j(m-j)(\ell^2_j +m\,\ell_j)\right. \\
&\left. +2 \sum_{j=2}^{m-1} \sum_{k<j} (m-j)k\,\ell_m\,\ell_k \right],
\end{split}
\end{equation}
where the Dynkin label $\ell_j$ is the number of columns  with $j$ boxes in a Young diagram. Obviously, $t_m(\overline{R})=t_m(R)$. The Dynkin index in Eq.~\eqref{eq:C1} is properly normalized such that it is equal to $1/2$ for the fundamental representation in $\mathsf{SU}(N)$ groups. In particular~\cite{Emmanuel-Costa:2013gia},
\begin{equation}\label{eq:t1t2R}
t_1(R)=y_R^2\,,\qquad t_2(R)=\,\frac{d_2(R)\left[d^{\;2}_2(R)-1\right]}{12}\,,
\end{equation}
for the subgroups $\mathsf{U}(1)_Y$ and $\mathsf{SU}(2)$, respectively; $y_R$ is the hypercharge for the representation~$R$. The corresponding Dynkin index $t_3(R)$ and cubic anomaly index $A_3(R)$ for $\mathsf{SU}(3)$ are presented in Table~\ref{tab:particle}. An important constraint on the Dynkin index $t_2(R)$ comes from Witten's anomaly condition, i.e., $\sum_R\,t_2(R)$ must be an integer number, in order to avoid the global $\mathsf{SU}(2)$ gauge anomaly~\cite{Witten:1982fp}. Restricting ourselves, for simplicity, to $\mathsf{SU}(2)$ representations up to dimension five ($d_2(R)\leq5$),  this condition implies that the number of Weyl fermion doublets is even.

\begin{table}[t]
\centering
\caption{\label{tab:particle} The Dynkin and cubic anomaly indices, $t_3$  and $A_3$, for the smallest irreducible representations of $\mathsf{SU}(3)$.}
\begin{ruledtabular}
\begin{tabular}{ccccccccccccccc}
$\mathsf{SU}(3)$-irrep & & $\mathbf{3}$ & & $\mathbf{6}$ & & $\mathbf{8}$ & & $\mathbf{10}$ & & $\mathbf{15}$ & & $\mathbf{15^{\prime}}$\\
\hline
$t_3$ & & $\frac{1}{2}$ & & $\frac{5}{2}$ & & 3 & & $\frac{15}{2}$& & $10$& & $\frac{35}{2}$\\
$A_3$ & & 1 & & 7 & & 0 & & 27 && 14 && 77 \\
\end{tabular}
\end{ruledtabular}
\end{table}

Notice that the set of equations given in Eqs.~\eqref{eq:anomaly} is invariant under an overall rescaling of the hypercharge $y_R$ of all multiplets; therefore one has to choose properly the overall normalization of the hypercharge in order to fulfil the requirement of having vectorlike particles after the electroweak symmetry breaking. Obviously, having only one extra chiral fermion would imply that it must belong to an adjoint representation with vanishing hypercharge. Furthermore, it is straightforward to demonstrate that, in the presence of only two extra multiplets, the conditions given in Eqs.~\eqref{eq:anomaly} and the requirement of having vectorlike particles after the electroweak symmetry breaking imply that the multiplets must necessarily form a vectorlike set; otherwise, they must be chiral fermions belonging to adjoint representations, both with zero hypercharge. Indeed, let us rewrite Eqs.~\eqref{eq:anomaly} for the case of two arbitrary multiplets,
\begin{subequations}
\label{eq:anomaly2m}
\begin{align}
\label{eq:anomaly2ma}
&A_3(R_1)\,d_2(R_1)\,+\,A_3(R_2)\,d_2(R_2) \,=\, 0\,,\\
\label{eq:anomaly2mb}
&y_{R_1}t_3(R_1)\,d_2(R_1)\,+\,y_{R_2}t_3(R_2)\,d_2(R_2) \,=\, 0\,,\\
\label{eq:anomaly2mc}
&y_{R_1}t_2(R_1)\,d_3(R_1)\,+\,y_{R_2}t_2(R_2)\,d_3(R_2) \,=\, 0\,,\\
\label{eq:anomaly2md}
&y^3_{R_1}d_2(R_1)\,d_3(R_1)\,+\,y^3_{R_2}d_2(R_2)\,d_3(R_2) \,=\, 0\,,\\
\label{eq:anomaly2me}
&y_{R_1}d_2(R_1)\,d_3(R_1)\,+\,y_{R_2}d_2(R_2)\,d_3(R_2) \,=\, 0\,.
\end{align}
\end{subequations}

From Eqs.~(\ref{eq:anomaly2md})-(\ref{eq:anomaly2me}), we immediately conclude that $y^2_{R_1}=y^2_{R_2}$. Therefore, there are just two possibilities. If $y_{R_1}=y_{R_2}=y_R$ then Eq.~(\ref{eq:anomaly2m}e) becomes
\begin{equation} y_R\left(d_2(R_1)\,d_3(R_1)\,+\,d_2(R_2)\,d_3(R_2)\right) \,=\, 0\,.
\end{equation} Since $d_i(R)>0$, the unique solution is $y_R=0$. This condition leads to two types of solutions, depending on how the remaining Eq.~(\ref{eq:anomaly2ma}) is satisfied. One possibility is that anomaly cancellation occurs between both multiplets, for instance, as in the multiplet combination $\mathbf{(6,1)}_{0}\oplus\mathbf{ (\overline{3},7)}_{0}$.\footnote{Hereafter, chiral multiplets are represented as $(d_3(R),d_2(R))_{y_R}$, following the standard notation.} The other case is when Eq.~(\ref{eq:anomaly2ma}) is verified independently for each multiplet [i.e., $A_3(R_1)=A_3(R_2)=0$], which is similar to the case of adding just one chiral fermion.

Instead, if $y_{R_1}=-y_{R_2} \neq 0$ then Eq.~(\ref{eq:anomaly2me}) becomes
\begin{equation}
d_2(R_1)\,d_3(R_1) \,=\, d_2(R_2)\,d_3(R_2)\,,
\end{equation}
which when substituted in Eq.~(\ref{eq:anomaly2mc}) implies
\begin{equation}
t_2(R_1)\,d_2(R_2) \,=\, t_2(R_2)\,d_2(R_1).
\end{equation}
Using Eq.~\eqref{eq:t1t2R}, we then conclude that $d_2(R_1)\,=\,d_2(R_2)$. In other words, the two multiplets must have the same $\mathsf{SU}(2)$ representation. One can now rewrite the remaining anomaly relations simply as
\begin{equation}
\dfrac{d_3(R_1)}{d_3(R_2)}\,=\,\dfrac{t_3(R_1)}{t_3(R_2)}\,= \,-\dfrac{A_3(R_1)}{A_3(R_2)}\,=\,1\,.
\end{equation}
If we now require that particles are vectorlike with respect to color then, by definition, the $\mathsf{SU}(3)$ representations of $R_1$ and $R_2$ must be conjugates of each other, unless they are adjoint. Combining these results, we obtain the vectorlike set of two multiplets
\begin{equation}
\mathbf{(d,d^\prime)}_{y}\oplus\mathbf{(\overline{d},d^\prime)}_{-y}\,.
\end{equation}

The question naturally arises whether it is possible to find anomaly-free sets with only three chiral multiplets. Considering $\mathsf{SU}(3)$ representations with dimensions $\mathbf{d}\equiv d_3(R)\leq10$ and $\mathsf{SU}(2)$ representations with $d_2(R)\leq5$, and restricting our search to hypercharges given by rational numbers, we obtain four different sets of solutions, which are presented in Table~\ref{tabMinimalanom}. The parameter $z$ in Table~\ref{tabMinimalanom} reflects the fact that the hypercharge for each chiral particle can be rescaled by an overall amount. Note that all multiplets in each minimal set have the same $\mathsf{SU}(3)$ dimension $\mathbf{d}$. This is so because we restrict our analysis to $d_3(R)\leq10$. If one relaxes this constraint, solutions with different $\mathsf{SU}(3)$ dimensions are found; for instance, the set $\mathbf{(15,1)}_{z/6} \oplus\mathbf{(\overline{6},2)}_{-z/3} \oplus \mathbf{(1,3)}_{z/2}$ is anomaly free. Furthermore, for the sets P1 and P4, Witten's anomaly condition forces $\mathbf{d}$ to be an even number.

If one requires that the three chiral multiplets in each set give rise to vectorlike particles (with respect to the electric charge) after the gauge symmetry breaking, the following condition must also be verified:
\begin{equation}
\label{eq:vl}
\sum^3_{p=1}\sum_{j_p} \bigl[j_p+y_p(z)\bigr]^m \,=\,0\,,
\end{equation}
for any odd positive integer $m$. In the above equation, $j_p=-s_p,-s_p+1,\ldots,s_p-1,s_p$ with $2s_p+1=d_2(R_p)$. This condition would then fix the value of $z$. For $m=1$ or $3$, Eq.~\eqref{eq:vl} is automatically satisfied for any value of $z$ by virtue of the anomaly cancellation conditions given in Eqs.~\eqref{eq:anomaly}. The value of $z$ is therefore determined by taking $m=5$. This leads to $|z| = 0,1$, or $3$, regardless of the chosen set. However, we must also ensure that the solutions are simultaneously vectorlike with respect to $\mathsf{SU}(3)$. Obviously, when $\mathbf{d}=1$ or $8$, this is already guaranteed. On the other hand, for $\mathbf{d}=3,\,6$, or $10$, one can easily show that this requirement imposes $|z|=1$.

\begin{table}[t]
\centering
\caption{\label{tabMinimalanom} Minimal anomaly-free chiral sets for an arbitrary $\mathsf{SU}(3)$ dimension $\mathbf{d} \leq \mathbf{10}$ and $\mathsf{SU}(2)$ dimension $d_2(R)\leq5$. For the sets P1 and P4, Witten's anomaly condition forces $\mathbf{d}$ to be an even number.}
\begin{ruledtabular}
\begin{tabular}{clclcl}
Set & \multicolumn{5}{c}{Particle content} \\
\hline
P1 & $(\mathbf{d},\mathbf{1})_{5z/6}$ & $\oplus$ & $(\mathbf{d},\mathbf{2})_{-2z/3}$ & $\oplus$ & $(\overline{\mathbf{d}},\mathbf{3})_{z/6}$\\
P2 & $(\mathbf{d},\mathbf{1})_{7z/6}$ & $\oplus$ & $(\mathbf{d},\mathbf{3})_{-5z/6}$ & $\oplus$ & $(\overline{\mathbf{d}},\mathbf{4})_{z/3}$\\
P3 & $(\mathbf{d},\mathbf{1})_{3z/2}$ & $\oplus$ & $(\mathbf{d},\mathbf{4})_{-z}$    & $\oplus$ & $(\overline{\mathbf{d}},\mathbf{5})_{z/2}$\\
P4 & $(\mathbf{d},\mathbf{2})_{4z/3}$ & $\oplus$ & $(\mathbf{d},\mathbf{3})_{-7z/6}$ & $\oplus$ & $(\overline{\mathbf{d}},\mathbf{5})_{z/6}$\\
\end{tabular}
\end{ruledtabular}
\end{table}

\subsection{Gauge coupling unification}
\label{subsec:GCU}

\begin{figure}[t]
\centering
\includegraphics[width=0.95\linewidth]{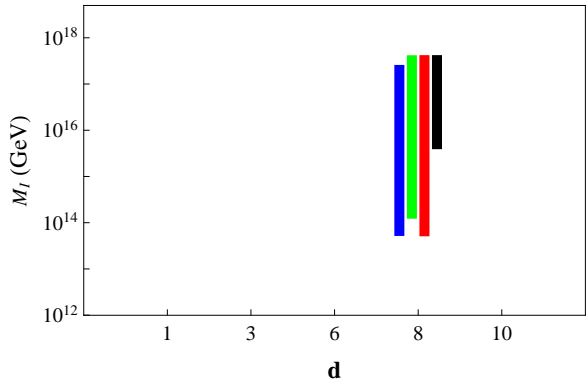}
\caption{\label{fig:P1a3} The allowed range of the intermediate scales $M_I$ and the unification scale $\Lambda$ as functions of the $\mathsf{SU}(3)$ dimension $\mathbf{d}$ for the solution P1 of Table~\ref{tabMinimalanom} with $z=3$. For each value of $\mathbf{d}$, the colored bars correspond (from left to right) to the energy scales $M_1$ $\bigl((\mathbf{d},\mathbf{1})_{5z/6}\bigr)$, $M_2$ $\bigl((\mathbf{d},\mathbf{2})_{-2z/3}\bigr)$, $M_3$ $\bigl((\overline{\mathbf{d}},\mathbf{3})_{z/6}\bigr)$, and $\Lambda$, respectively. }
\end{figure}

Let us now study the possibility of achieving gauge coupling unification with the anomaly-free sets previously found. We shall work in the framework of the SM, extended by new anomaly-free chiral sets of fermions. In our approach, we do not modify the Higgs and gauge sectors of the SM. More precisely, we assume that if an additional Higgs or gauge particle content beyond the SM is present, it does not affect the evolution of the running gauge couplings. Clearly, the new fermions would have to acquire masses by some mechanism that relies on an extended theory. A simple possibility is to extend the scalar sector and construct renormalizable Yukawa terms for the new chiral fields. Let us consider, for instance, the set P1 of Table~\ref{tabMinimalanom}, with $\psi_1\sim(\mathbf{8},\mathbf{1})_{5/2}$, $\psi_2\sim(\mathbf{8},\mathbf{2})_{-2}$, $\psi_3\sim(\mathbf{8},\mathbf{3})_{1/2}$. We then introduce the scalar fields $H\sim(\mathbf{1},\mathbf{2})_{-1/2}$, $\Phi\sim(\mathbf{1},\mathbf{4})_{3/2}$, $\Delta\sim(\mathbf{1},\mathbf{3})_{-1}$ and write the Yukawa interaction terms $\epsilon_{ab}\psi_1\psi^{a}_2H^{b} \,$, $\epsilon_{ad}\epsilon_{be}\epsilon_{cf}\psi^{a}_2\psi^{bc}_3\Phi^{def}$ and $\epsilon_{ac}\epsilon_{be}\epsilon_{df}\psi^{ab}_3\psi^{cd}_3\Delta^{ef}\,$. The new fermions become massive once the neutral components of the scalar fields acquire vacuum expectation values, i.e., $\langle H^1\rangle\neq0$, $\langle\Phi^{222}\rangle\neq0$, and $\langle\Delta^{11}\rangle\neq0$. The limitation of this mechanism is that the masses of the new physical fermions are tightly constrained by electroweak precision data. Indeed, since the new scalar fields should carry nontrivial charges under the SM group, they also contribute to the electroweak gauge boson masses. As it turns out, the inclusion of such scalars leads to gauge coupling unification only if some of their masses are much higher than the electroweak scale.  Moreover, for the cases where unification is achieved, the masses of the new fermions cannot be all lowered to the electroweak scale. Thus, this mass mechanism is disfavored in our minimal framework. 

If we insist on the unification of the gauge couplings at a high energy scale, we are led to consider an alternative mechanism that allows for a successful unification, without imposing tight constraints on the masses of the new fermionic degrees of freedom. In such a case, these extra particles could decouple from the theory at intermediate energy scales. An attractive possibility is to enlarge the fermionic sector by considering mirror fermions~\cite{Montvay:1987ys}, as in chiral gauge theories based on noncommutative geometry~\cite{Connes:1994yd}. The advantage of this approach is that it can provide a dynamical mechanism in which fermions belonging to the physical subspace acquire masses at intermediate scales while their mirror partner states get masses higher than the unification scale~\cite{Lizzi:1996vr,Lizzi:1997sg}. Obviously, a detailed analysis would be required to fully implement this paradigm in our framework. In what follows, we shall assume that whatever mechanism is chosen, it allows for the decoupling of some chiral fermions at intermediate mass scales, much higher than the electroweak scale.

\begin{figure}[t]
\centering
 \includegraphics[width=0.95\linewidth]{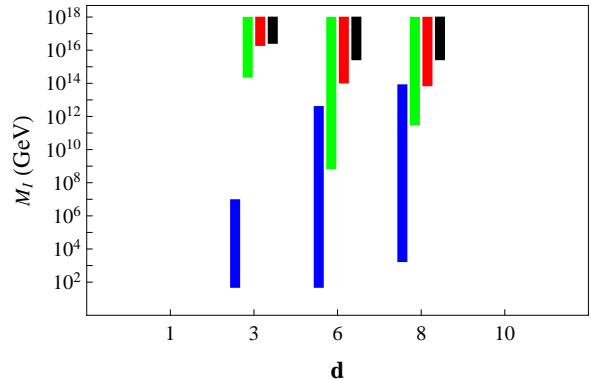}
\hfill \includegraphics[width=0.95\linewidth]{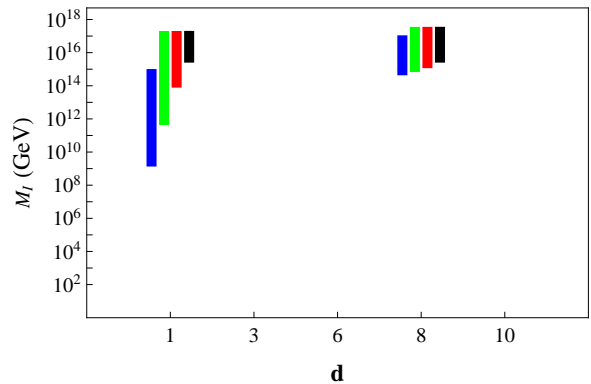}
\caption{\label{fig:P2}
The allowed range of the intermediate scales $M_I$ and the unification scale $\Lambda$ as functions of the $\mathsf{SU}(3)$ dimension $\mathbf{d}$ for the solution P2 of Table~\ref{tabMinimalanom} with $z=1$~(upper panel) and $z=3$~(lower panel). For each value of $\mathbf{d}$, the colored bars correspond (from left to right) to the energy scales $M_1$ $\bigl((\mathbf{d},\mathbf{1})_{7z/6}\bigr)$, $M_2$ $\bigl((\mathbf{d},\mathbf{3})_{-5z/6}\bigr)$, $M_3$ $\bigl((\overline{\mathbf{d}},\mathbf{4})_{z/3}\bigr)$, and $\Lambda$, respectively.}
\end{figure}

\begin{figure}[t]
	\centering
	\includegraphics[width=0.95\linewidth]{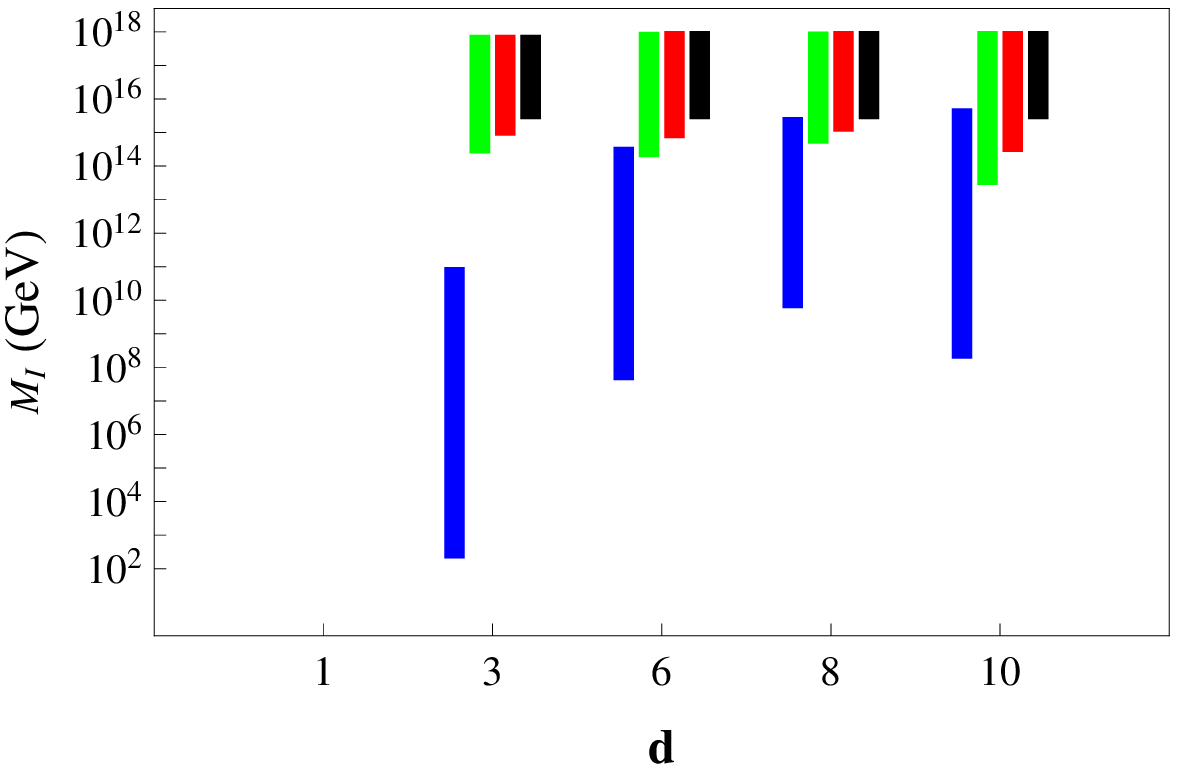}
	\hfill\includegraphics[width=0.95\linewidth]{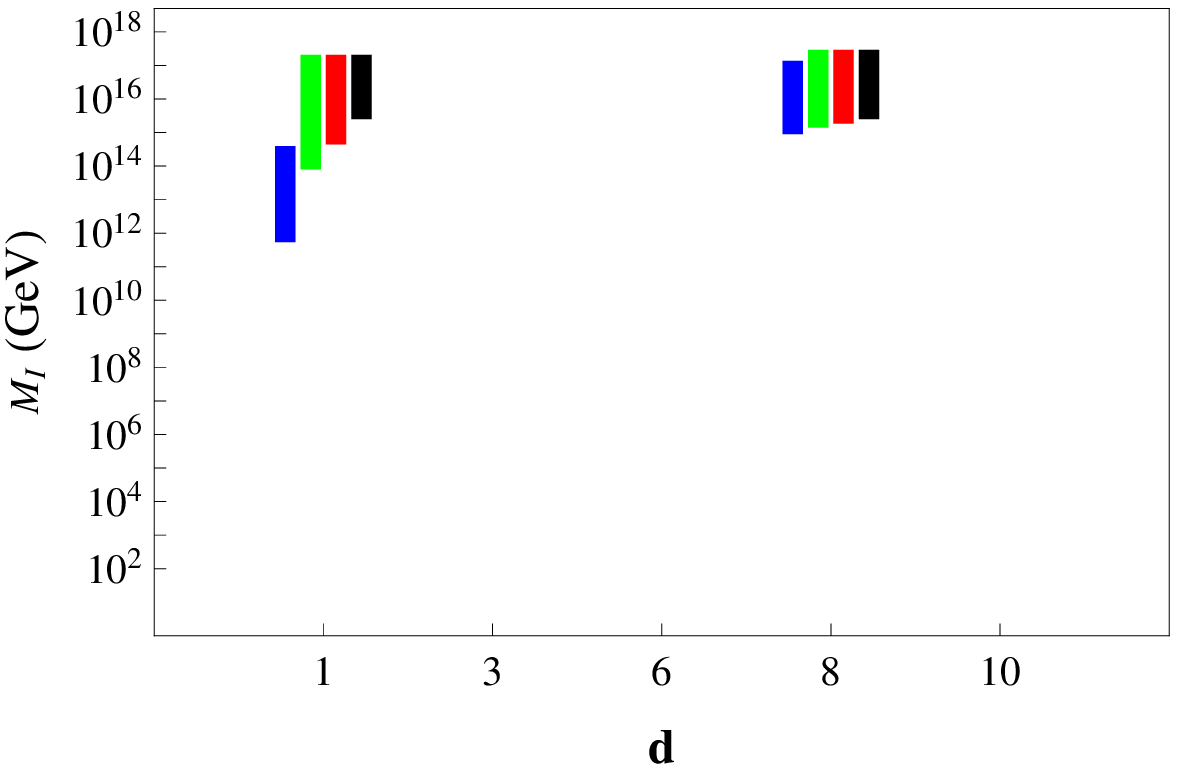}
	\caption{\label{fig:P3} The allowed range of the intermediate scales $M_I$ and the unification scale $\Lambda$ as functions of the $\mathsf{SU}(3)$ dimension $\mathbf{d}$ for the solution P3 of Table~\ref{tabMinimalanom} with $z=1$~(upper panel) and $z=3$~(lower panel). For each value of $\mathbf{d}$, the colored bars correspond (from left to right) to the energy scales $M_1$ $\bigl((\mathbf{d},\mathbf{1})_{3z/2}\bigr)$, $M_2$ $\bigl((\mathbf{d},\mathbf{4})_{-z}\bigr)$, $M_3$ $\bigl((\overline{\mathbf{d}},\mathbf{5})_{z/2}\bigr)$, and $\Lambda$, respectively.}
\end{figure}

\begin{figure}[t]
	\centering
	\includegraphics[width=0.95\linewidth]{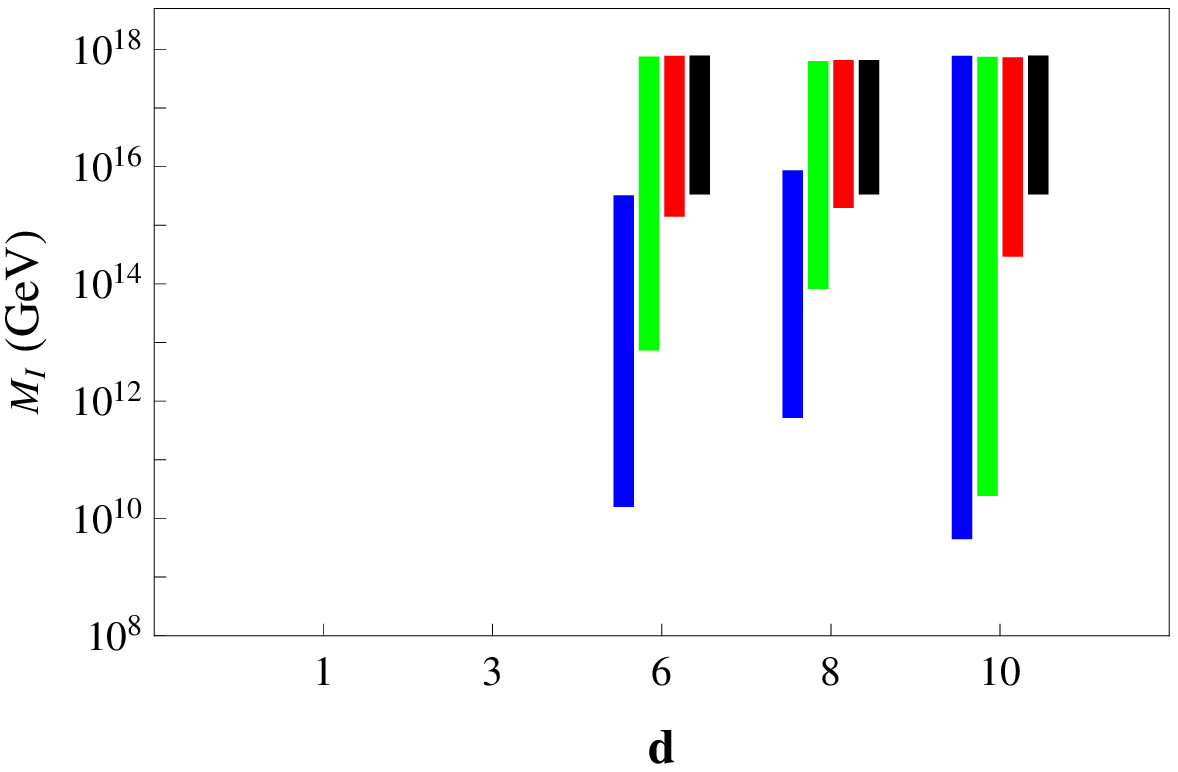}
	\hfill\includegraphics[width=0.95\linewidth]{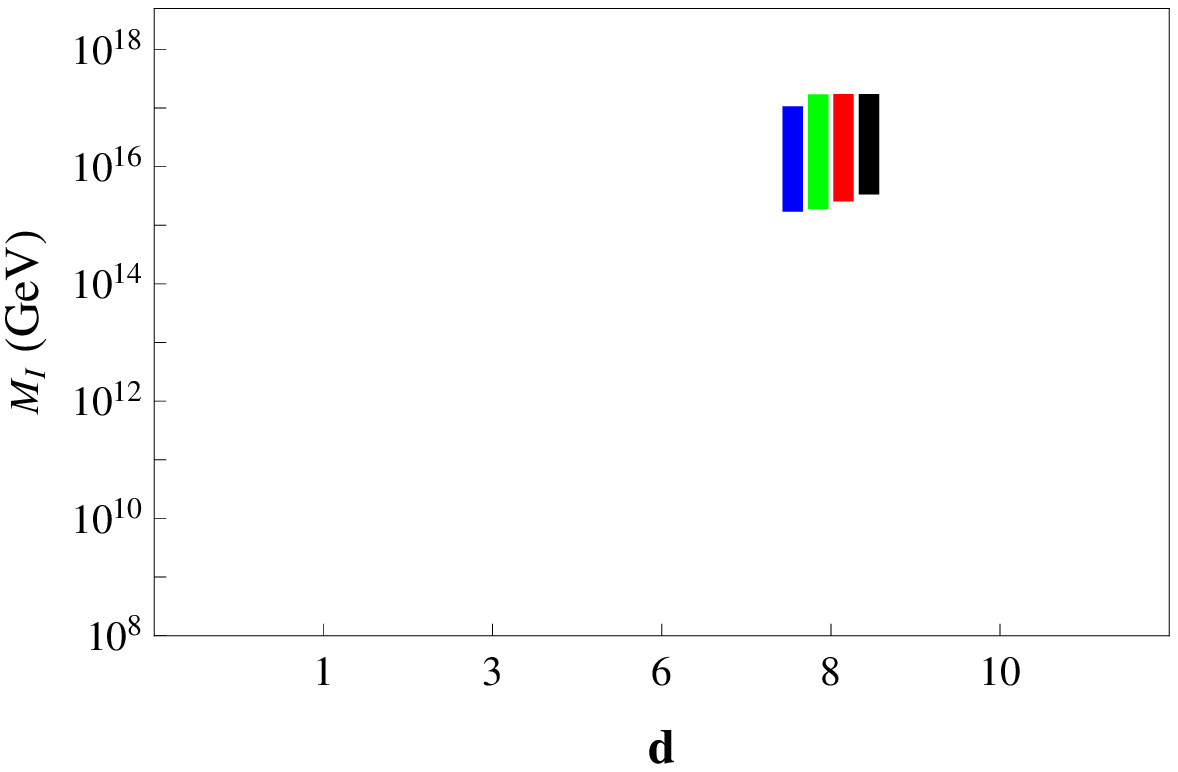}
	
	\caption{\label{fig:P4} The allowed range of the intermediate scales $M_I$ and the unification scale $\Lambda$ as functions of the $\mathsf{SU}(3)$ dimension $\mathbf{d}$ for the solution P4 of Table~\ref{tabMinimalanom} with $z=1$~(upper panel) and $z=3$~(lower panel). For each value of $\mathbf{d}$, the colored bars correspond (from left to right) to the energy scales $M_1$ $\bigl((\mathbf{d},\mathbf{2})_{4z/3}\bigr)$, $M_2$ $\bigl((\mathbf{d},\mathbf{3})_{-7z/6}\bigr)$, $M_3$ $\bigl((\overline{\mathbf{d}},\mathbf{5})_{z/6}\bigr)$, and $\Lambda$, respectively.}
\end{figure}

The evolution of gauge couplings $\alpha_i\, (i=1,2,3)$ is governed by the RGEs. Assuming the presence of $N$ chiral fermions with intermediate mass scales $M_I,I=1,\ldots,N$, and  $M_Z\leq M_I\leq\Lambda$, the one-loop solution at the unification scale $\Lambda$ can be written as
\begin{equation}
\label{eq:iasolution}
\alpha^{-1}_i(M_Z)=\alpha^{-1}_{U}+\frac{1}{2\pi}B_i\ln\biggl(\frac{\Lambda}{M_Z}\biggr)\,,
\end{equation}
with
\begin{equation}
\label{eq:unifgut}
\alpha_U\equiv\kappa_1\,\alpha_y(\Lambda)=\kappa_2\,\alpha_w(\Lambda)=\kappa_3\,\alpha_s(\Lambda)\,,
\end{equation}
and $\alpha_U \lesssim 1$ to ensure that the perturbative regime holds~\cite{Kopp:2009xt}. In Eq.~\eqref{eq:iasolution},
\begin{equation}
\label{eq:Bi}
B_i=\frac{1}{\kappa_i}\biggl(b_i + \sum^N_{I=1}  b_i^I r_I \biggr)\,,
\end{equation}
where the ``running weight'' $r_I$ is defined for each intermediate energy scale $M_I$ as
\begin{equation}
\label{eq:ri}
r_I=\frac{\ln\left(\Lambda/M_I\right)}{\ln\left(\Lambda/M_Z\right)}\,,
\end{equation}
and takes values in the interval $0\leq r_I \leq 1$. The one-loop beta coefficients $b_i$ account for the SM contribution, while $b^I_i$ are the contributions of the intermediate particles above the threshold $M_I$. For a given particle in a representation $R$, the one-loop beta coefficients $b^R_i$ are computed using the formula
\begin{equation}
\label{eq:bi}
b^R_i= s(R)\, t_i(R)\prod_{j\neq i}d_j(R)\,,
\end{equation}
where $s(R)=1/6$ for a real scalar,  $1/3$ for a complex scalar, $-11/3$ for a gauge boson, $2/3$ for a chiral fermion, and $4/3$ for a vectorlike fermion. When applied to the SM particle content, this gives $b_1=41/6$, $b_2=-19/6$, and $b_3=-7$.

In order to verify whether the minimal anomaly-free sets in Table~\ref{tabMinimalanom} lead to the unification of the gauge couplings, we need to solve Eqs.~\eqref{eq:unifgut}. Each set of solutions in Table~\ref{tabMinimalanom} is composed of $N=3$ chiral multiplets. Thus, there are two equations to determine four unknowns: the intermediate mass scales of the three extra particles, given by the running weights $r_{1,2,3}$, and the unification scale $\Lambda$.

We choose to vary the running weights $r_2$ and $r_3$ in the allowed range $[0,1]$, and then determine the scale of the first particle in the set,
\begin{equation}
\label{eq:r1r2GUT}
r_1=\frac{B\,B^{\prime}_{12}-B^{\prime}_{23}}{\Delta^1_{23}- B \Delta^1_{12}}\,,
\end{equation}
where
\begin{equation}
\label{eq:Bidif}
B^{\prime}_{ij}= B^{\prime}_i - B^{\prime}_j\,,\quad
B^{\prime}_i=\frac{1}{\kappa_i}\biggl(b_i \, + \, \sum^N_{I=2} b^I_{i} r_I\biggr)\,,
\end{equation}
and
\begin{equation}
\Delta^I_{ij}=\frac{b^I_i}{\kappa_i}-\frac{b^I_j}{\kappa_j}\,.
\end{equation}
The unification scale $\Lambda$ is obtained through the relation
\begin{equation}
\label{eq:unifscale}
\ln \biggl(\frac{\Lambda}{M_Z}\biggr)=\frac{\widetilde{B}}{B_{1}-B_{2}}\,.
\end{equation}

The constants $B$ and $\widetilde{B}$, defined in terms of parameters at $M_Z$ scale, are given by~\cite{Giveon:1991zm,EmmanuelCosta:2011wp}
\begin{equation}
\label{eq:B}
\begin{split}
B &\equiv \frac{\sin^2 \theta_w - \dfrac{\kappa_2\,\alpha}{\kappa_3\,\alpha_s}}{\dfrac{\kappa_2}{\kappa_1} -\left(1+\dfrac{\kappa_2}{\kappa_1}\right)\sin^2 \theta_w }\,,\\
\widetilde{B}&\equiv \frac{2\pi}{\alpha}\left[\frac{1}{\kappa_1}-\left(\frac{1}{\kappa_1} +\frac{1}{\kappa_2}\right)\sin^2\theta_w \right].
\end{split}
\end{equation}
Using the experimental data at the electroweak scale, chosen here at  $M_Z=91.1876\text{ GeV}$~\cite{Beringer:1900zz},
\begin{equation}
\label{eq:alphas}
\begin{aligned}
\alpha^{-1}&=127.944\pm 0.014\,,\\
\alpha_s &= 0.1185\pm 0.0006\,,\\
\sin^2 \theta_w &= 0.23126\pm 0.00005\,,
\end{aligned}
\end{equation}
and considering $\kappa_1=\kappa_2=\kappa_3=1$ we obtain
\begin{equation}
\label{eq:actualB4K111}
B=0.308\pm0.001, \quad  \widetilde{B}= 431.4 \pm 0.1\,.
\end{equation}
If one uses instead the $\mathsf{SU}(5)$ normalization, then
\begin{equation}
\label{eq:actualB}
B=0.718\pm0.003,\quad \widetilde{B}=185.0\pm0.2\,.
\end{equation}

In Figs.~\ref{fig:P1a3}-\ref{fig:P4}, we present the allowed ranges of the intermediate scales $M_I$ and the unification scale $\Lambda$ as functions of the $\mathsf{SU}(3)$ dimension $\mathbf{d}$ for the 16 solutions found within the chiral sets of Table~\ref{tabMinimalanom}. In all cases, we require $\Lambda \geq 5\times10^{15}$~GeV, so that we are safely consistent with proton decay bounds. First, we note that no solution with $z=0$, i.e., with vanishing hypercharge, and $\mathbf{d}=1$ or $8$, leads to successful unification. For set P1, gauge coupling unification is attained only for the hypercharge normalization $z=3$ and for $\mathbf{d}=8$. The required intermediate scales turn out to be above $10^{13}$~GeV. In the case of set P2, when $z=1$, solutions were found only with $\mathbf{d}=3,\, 6$, or $8$. In this case, the mass of the $\mathsf{SU}(2)$ singlet, $M_1$, can take values at the TeV scale or even lower, while the other two masses are above $10^{9}$~GeV. When $z=3$, solutions were found for the allowed values of $\mathbf{d}=1$ or $8$. We note that for $\mathbf{d}=8$ unification is achieved for quite high intermediate scales, between $10^{14}$~GeV and the unification scale. The mass and unification behavior in the set P3 is quite similar to the set P2. In this case, when $z=1$, solutions were found also for $\mathbf{d}=10$. Finally, the set P4 unifies for $z=1$ with any $\mathsf{SU}(3)$ even dimension, while for $z=3$ solutions were found with $\mathbf{d}=8$. The intermediate scales in this set have values above $10^{10}$~and $10^{15}$~GeV, for $z=1$ and $z=3$, respectively.

Let us now analyze whether the above solutions also lead to the unification of gauge and gravitational couplings at the string scale. In this case, in addition to Eqs.~\eqref{eq:iasolution}, the unification scale $\Lambda$ must also satisfy the constraint
\begin{equation}
\label{eq:astring}
\alpha_{U}\,=\,\alpha_{\text{string}}\,= \,\frac{1}{4\pi}\biggl(\frac{\Lambda}{\Lambda_S}\biggr)^2\,,
\end{equation}
where $\Lambda_{S} = 5.27\times 10^{17}\text{ GeV}$ is the string scale that takes into account one-loop string effects in the weak coupling limit~\cite{Kaplunovsky:1987rp}.

Since we have now three equations for unification, namely Eqs.~\eqref{eq:unifgut} and~\eqref{eq:astring}, we choose to determine the running weights of the first two particles of each set, $r_1$ and $r_2$, and the unification scale $\Lambda$, while $r_3$ is free to take values within the allowed range $[0,1]$. We find
\begin{equation}
\label{eq:ristring}
  r_1\,=\,\beta_2-\frac{\widetilde{\beta}_2}{\ln\left(\Lambda/M_Z\right)},
\quad
  r_2\,=\,-\beta_1+\frac{\widetilde{\beta}_1}{\ln\left(\Lambda/M_Z\right)}\,.
\end{equation}
The coefficients $\beta_i$ and $\widetilde{\beta}_i$ are given by
\begin{equation}
\begin{split}
\beta_i &=\,\frac{B^{\prime\prime}_{12}\,\Delta_{23}^i\, - \,B^{\prime\prime}_{23}\Delta_{12}^i}{\Delta_{12}^2\,\Delta_{23}^1\, -\,\Delta_{23}^2\,\Delta_{12}^1} \,,\\
\widetilde{\beta}_i &=\,\widetilde{B}\,\frac{  \Delta_{23}^i-B\,\Delta_{12}^i}{\Delta_{12}^2\,\Delta_{23}^1\, -\,\Delta_{23}^2\,\Delta_{12}^1}\,,
\end{split}
\end{equation}
where
\begin{equation}
\label{eq:Bidif2}
B^{\prime\prime}_{ij}= B^{\prime\prime}_i - B^{\prime\prime}_j\,,\quad
B^{\prime\prime}_i=\frac{1}{\kappa_i}\biggl(b_i \, + \, \sum^N_{I=3} b^I_{i} r_I\biggr)\,.
\end{equation}
The unification scale $\Lambda$ is then given by
\begin{equation}
\label{eq:lamber}
\biggl(\dfrac{\Lambda_S}{\Lambda}\biggr)^2 \,=\, -c\,{\rm W}_{k}
\left(-\frac{\Lambda^2_S}{c\,M_Z^2}\,e^{-\frac{a}{4\pi c}}\right)\,,
\end{equation}
where
\begin{equation}
\begin{split}
a\,&=\,\alpha^{-1}_{i}(M_Z)+\frac{1}{2\pi}\left(\widetilde{\beta}_2 \frac{b^1_i}{\kappa_i}-\widetilde{\beta}_1 \frac{b^2_i}{\kappa_i}\right)\,,\\
c\,&=\,\frac1{16\pi^2}\left(\beta_2 \frac{b^1_i}{\kappa_i}-\,\beta_1  \frac{b^2_i}{\kappa_i}\,+\,B^{\prime\prime}_i\right)\,,
\end{split}
\end{equation}
for any choice $i=1,2,3$. The function $W_k(x)$  is the Lambert W function, with  $k=0$ for $c<0$ and $k=-1$ for $c>0$. One can verify that the contribution of the running weight $r_3$ cancels out in the constants $a$ and $c$. Thus, the unification scale is uniquely determined for any given set of chiral multiplets.

\begin{table*}
\centering
\caption{\label{tab:paradigmaString} Minimal chiral sets of solutions for which there is string-scale unification. The particle content of each set is defined in Table~\ref{tabMinimalanom}. The mass ordering $M_1$, $M_2$, and $M_3$ corresponds to the ordering of the multiplets adopted for each set in Table~\ref{tabMinimalanom}.}
\begin{ruledtabular}
\begin{tabular}{ccccccc}
Set & $z$ & $\mathbf{d}$ & $\Lambda$ [$10^{17}$ GeV] & \textit{$M_1$} [GeV] & \textit{$M_2$} [GeV] & \textit{$M_3$} [GeV] \\
\hline
P1  & 3 & 8 & 2.7 & $[2.7,7.3]\times 10^{16}$ & $[1.2,2.7]\times 10^{17}$ & $[2.2,2.7]\times 10^{17}$ \\
 \\ 
\multirow{3}{*}{P2} & 1 & 3 & 2.7 & $[2.1,26]\times 10^{4}$ & $[4.4,26]\times 10^{16}$ & $[1.3,2.7]\times 10^{17}$ \\
 & 1 & 6 & 3.0 & $[3.7,45]\times 10^{10}$ & $[6.7,40]\times 10^{14}$ & $[1.4,3.0]\times 10^{17}$ \\
 & 1 & 8 & 2.9 & $[9.4,110]\times 10^{11}$ & $[8.7,52]\times 10^{15}$ & $[1.4,2.9]\times 10^{17}$ \\
 \\ 
\multirow{4}{*}{P3} & 1 & 3 & 2.7 & $[8.6,12]\times 10^{9}$ & $[2.2,2.7]\times 10^{17}$ & $[2.5,2.7]\times 10^{17}$ \\
 & 1 & 6 & 2.8 & $[1.3,4.4]\times 10^{13}$ & $[1.2,2.8]\times 10^{17}$ & $[1.9,2.8]\times 10^{17}$ \\
 & 1 & 8 & 2.8 & $[1.9,4.1]\times 10^{14}$ & $[1.7,2.8]\times 10^{17}$ & $[2.2,2.8]\times 10^{17}$ \\
 & 1 & 10 & 3.0 & $[1.1,9.2]\times 10^{14}$ & $[4.7,20]\times 10^{16}$ & $[1.5,3.0]\times 10^{17}$ \\
 \\ 
\multirow{3}{*}{P4} & 1 & 6 & 2.8 & $[3.5,13]\times 10^{14}$ & $[9.1,28]\times 10^{16}$ & $[2.4,2.8]\times 10^{17}$ \\
 & 1 & 8 & 2.8 & $[2.2,3.6]\times 10^{15}$ & $[1.8,2.8]\times 10^{17}$ & $[2.6,2.8]\times 10^{17}$ \\
 & 1 & 10 & 3.2 & $[3.1,320]\times 10^{15}$ & $[8.3,490]\times 10^{14}$ & $[1.6,2.8]\times 10^{17}$\\
\end{tabular}
\end{ruledtabular}
\end{table*}

Out of the 16 solutions, only 11 are consistent with string-scale unification. The results are summarized in Table~\ref{tab:paradigmaString}. In all cases, the unification scale $\Lambda$ is in the range $[2.7,3.2]\times 10^{17}$~GeV, which implies $34\leq\alpha^{-1}_{\text{string}}\leq48$. The intermediate mass $M_3$ is very close to the unification scale for all solutions, and no solutions were found with intermediate scales below the TeV scale. Moreover, solutions with the hypercharge normalization  $z=3$ were obtained only for set P1 with $\mathbf{d}=8$.

We conclude this section by noticing that the minimal anomaly-free sets given in Table~\ref{tabMinimalanom}, although leading to unification with only three chiral fermion multiplets, do not exhibit the usual quantum numbers under the SM group, such as those present in SM GUT embeddings into $\mathsf{SU}(5)$ or $\mathsf{SO}(10)$ groups.

\section{$\mathsf{SU}(5)$-inspired anomaly-free chiral fermion sets}
\label{sec:SU5}

\begin{table*}
\centering
\caption{\label{tab:ParticleBis} Particle multiplets and one-loop beta coefficients for the $\mathsf{SU}(5)$ representations with dimensions less than or equal to $50$.}

\begin{ruledtabular}
\begin{tabular}{clccclcc}
Label & Multiplet & $\mathsf{SU}(5)$-rep & $(b_1,b_2,b_3)$ & Label & Multiplet & $\mathsf{SU}(5)$-rep &
$(b_1,b_2,b_3)$\\
\hline
 1  & $\mathbf{(1,2)}_{1/2}$     &     $\mathsf{5},\,\mathsf{45}$   &   $(1/3,1/3,0)$    &
 12 & $\mathbf{(1,2)}_{-3/2}$   &   $\mathsf{40}$    &   $(3,1/3,0)$   \\
 \\
 2  & $\mathbf{(3,1)}_{-1/3}$    &   $\mathsf{5},\,\mathsf{45},\,\mathsf{50}$   &   $(2/9,0,1/3)$   &
 13 & $\mathbf{(8,1)}_{1}$       &    $\mathsf{40}$   &   $(16/3,0,2)$  \\
\\
 3  & $\mathbf{(1,1)}_{1}$         &   $\mathsf{10}$   &   $(2/3,0,0)$     &
 14 & $\mathbf{(3,3)}_{-1/3}$    &   $\mathsf{45}$    &   $(2/3,4,1)$   \\
 \\
 4  & $\mathbf{(\overline{3},1)}_{-2/3}$  & $\mathsf{10},\,\mathsf{40}$  &   $(8/9,0,1/3)$   &
 15 & $\mathbf{(\overline{3},1)}_{4/3}$  &  $\mathsf{45}$&   $(32/9,0,1/3)$  \\
 \\
 5  & $\mathbf{(3,2)}_{1/6}$         &    $\mathsf{10},\,\mathsf{15},\,\mathsf{40}$  &   $(1/9,1,2/3)$   &
 16 & $\mathbf{(\overline{3},2)}_{-7/6}$   &   $\mathsf{45},\,\mathsf{50}$ & $(49/9,1,2/3)$  \\
 \\
 6  & $\mathbf{(1,3)}_{1}$           &  $\mathsf{15}$    &   $(2,4/3,0)$     &
17 & $\mathbf{(\overline{6},1)}_{-1/3}$   &   $\mathsf{45}$ & $(4/9,0,5/3)$   \\
\\
7  & $\mathbf{(6,1)}_{-2/3}$         &   $\mathsf{15}$  &   $(16/9,0,5/3)$  &
 18 & $\mathbf{(8,2)}_{1/2}$         &   $\mathsf{45},\,\mathsf{50}$   &   $(8/3,8/3,4)$   \\
 \\
 8  & $\mathbf{(1,4)}_{-3/2}$        &    $\mathsf{35}$  &   $(6,10/3,0)$    &
 19 & $\mathbf{(1,1)}_{-2}$          &    $\mathsf{50}$  &   $(8/3,0,0)$     \\
 \\
 9  & $\mathbf{(\overline{3},3)}_{-2/3}$ & $\mathsf{35},\,\mathsf{40}$ &   $(8/3,4,1)$     &
   20 & $\mathbf{(\overline{6},3)}_{-1/3}$ &  $\mathsf{50}$ &   $(4/3,8,5)$     \\
\\
 10 & $\mathbf{(\overline{6},2)}_{1/6}$   & $\mathsf{35},\,\mathsf{40}$ &   $(2/9,2,10/3)$  &
 21 & $\mathbf{(6,1)}_{4/3}$      &  $\mathsf{50}$ &   $(64/9,0,5/3)$  \\
 \\
 11 & $\mathbf{(\overline{10},1)}_{1}$   &  $\mathsf{35}$   &   $(20/3,0,5)$   \\
\end{tabular}
\end{ruledtabular}
\end{table*}

In view of our previous results, and inspired by simplicity in $\mathsf{SU}(5)$ as a gauge group, we may ask what  the minimal chiral sets are, besides the SM combination $\overline{\mathsf{5}}\oplus\mathsf{10}$,  that fulfil our three requirements, namely, be anomaly free, vectorlike with respect to the color and electric charges, and, finally, lead to gauge unification.

We shall only consider $\mathsf{SU}(5)$ representations with dimensions less than or equal to 50. For the $\mathsf{SU}(5)$ multiplets contained in these representations~\cite{Slansky:1981yr}, we give in Table~\ref{tab:ParticleBis} the corresponding one-loop beta coefficients $b_i$. The representations are labeled from 1 to 21 according to their quantum numbers.

Applying the anomaly constraints in Eqs.~\eqref{eq:anomaly} to the 21 particle species of Table~\ref{tab:ParticleBis}, we obtain the following system of linear equations for the number of multiplets, $n_I,\, I=1, \ldots, 21$, of each particle type:
\begin{widetext}
\begin{equation}\label{eq:anomalySU5}
\begin{aligned}
n_{2} &- n_{4} + 2n_{5} + 7n_{7} - 3n_{9} - 14n_{10} - 27n_{11} + 3n_{14} - n_{15} - 2n_{16}- 7n_{17} - 21n_{20} + 7n_{21}=0\,,   
\\ \\
n_{2} &+ 2n_{4} - n_{5} + 10n_{7} + 6n_{9} - 5n_{10} - 45n_{11} - 18n_{13} + 3n_{14} - 4n_{15}+ 7n_{16} + 5n_{17}\\
&   - 18n_{18} +15n_{20} - 20n_{21}=0\,,   
\\ \\
n_{1}  &  + n_{5} + 8n_{6}  -30n_{8} -16n_{9} + 2n_{10}  -3n_{12}  -8n_{14}  -7n_{16}  + 8n_{18}  -16n_{20}  =0\,,   
\\ \\
9n_{1}& -4n_{2} + 36n_{3} -32n_{4} + n_{5} + 108n_{6} -64n_{7} -486n_{8} -96n_{9} + 2n_{10} + 360n_{11} -243n_{12}  \\
&+ 288n_{13} -12n_{14} + 256n_{15} -343n_{16} -8n_{17} + 72n_{18} -288n_{19} -24n_{20} + 512n_{21} =0\,,   
\\ \\
n_{1}  & -n_{2} + n_{3} -2n_{4} + n_{5} + 3n_{6} -4n_{7} -6n_{8} -6n_{9} + 2n_{10} + 10n_{11} -3n_{12} + 8n_{13} - 3n_{14} \\
&+ 4n_{15} -7n_{16} -2n_{17} + 8n_{18} -2n_{19} -6n_{20} + 8n_{21} =0\,. 
\end{aligned}
\end{equation}
\end{widetext}

Besides the anomaly constraints, we must also require that the new low-energy fermion states form vectorlike sets with respect to the color $\mathsf{SU}(3)$ and electromagnetic $\mathsf{U}(1)$. This requirement leads to the additional constraints
\begin{equation}
\label{eq:vectorlikeSU5}
\begin{aligned}
\mathbf{1}_{1}: & \quad n_{1}\,+\,n_{3}\,+\,n_{6}\,-\,n_{8}\,-\,n_{12}\,=\,0\,,\\
\mathbf{3}_{-1/3}: & \quad n_{2}\,+\,n_{5}\,-\,n_{9}\,+\,n_{14}\,=\,0\,,\\
\overline{\mathbf{3}}_{-2/3}: & \quad n_{4}\,-\,n_{5}\,+\,n_{9}\,-\,n_{14}\,+\,n_{16}\,=\,0\,,\\
\mathbf{1}_{2}: & \quad n_{6}\,-\,n_{8}\,-\,n_{12}\,-\,n_{19}\,=\,0\,,\\
\mathbf{6}_{-2/3}: & \quad n_{7}\,-\,n_{10}\,-\,n_{20}\,=\,0\,,\\
\mathbf{1}_{-3}: & \quad n_{8}\,=\,0\,,\\
\overline{\mathbf{3}}_{-5/3}: & \quad n_{9}\,+\,n_{16}\,=\,0\,,\\
\overline{\mathbf{6}}_{-1/3}: & \quad n_{10}\,+\,n_{17}\,+\,n_{20}\,=\,0\,,\\
\overline{\mathbf{10}}_{1}: & \quad n_{11}\,=\,0\,,\\
\mathbf{8}_{1}: & \quad n_{13}\,+\,n_{18}\,=\,0\,,\\
\mathbf{3}_{-4/3}: & \quad n_{14}\,-\,n_{15}\,=\,0\,,\\
\overline{\mathbf{6}}_{-4/3}: & \quad n_{20}\,-\,n_{21}\,=\,0\,,
\end{aligned}
\end{equation}
where $\mathbf{d}_Q$ denotes states with $\mathsf{SU}(3)$ dimension $\mathbf{d}$ and electric charge $Q$. Substituting Eqs.~\eqref{eq:vectorlikeSU5} into Eqs.~\eqref{eq:anomalySU5}, we verify that the first, second, and last equations in~\eqref{eq:anomalySU5} are automatically satisfied, while the remaining two equations can be rewritten as
\begin{equation}
\label{eq:combinedSU5}
\begin{aligned}
&2n_1 + 9n_2 + 3n_3 + 17n_4 - 9n_5 - 5n_6 + 16n_7 \\
&- 18n_{10} + 8n_{13} = 0\,, \\
&54n_1  +  243n_2 +  81n_3 + 459n_4  - 243n_5 - 135n_6 \\
&+ 432n_7 - 486n_{10} + 216n_{13}=0\,.
\end{aligned}
\end{equation}

\begin{table*}
	\centering
	\caption{\label{tab:Content} Anomaly-free chiral sets obtained from $\mathsf{SU}(5)$ representations up to dimension 50. The multiplet notation follows the convention $(\mathsf{SU}(3),\mathsf{SU}(2))_{Y}$, and $n_s$ corresponds to the number of different species required in each set. For each solution it is also indicated whether it leads to GUT (G) and/or string (S) scale unification. The sets S9, S19, and S20, marked with $\ast$, unify at a scale $\Lambda\lesssim10^ {16}$ GeV.}
	\begin{ruledtabular}
		\begin{tabular}{cclcc}
			Set & $n_s$ & \multicolumn{1}{c}{Particle content}  & G & S\\
			\hline
			S1  & 4  & $3\,\mathbf{(1,2)}_{1/2} \oplus 2\,\mathbf{(1,1)}_{-1} \oplus \mathbf{(1,2)}_{-3/2} \oplus \mathbf{(1,1)}_{2}$    & -- & -- \\
			S2  & 5  & $\mathbf{(1,2)}_{1/2} \oplus \mathbf{(3,1)}_{-1/3} \oplus \mathbf{(1,1)}_{-1} \oplus \mathbf{(3,1)}_{2/3} \oplus \mathbf{(\overline{3},2)}_{-1/6}$  & -- & -- \\
			S3  & 5  & $\mathbf{(1,1)}_{-1} \oplus \mathbf{(1,3)}_{1} \oplus \mathbf{(8,1)}_{1} \oplus \mathbf{(8,2)}_{-1/2} \oplus \mathbf{(1,1)}_{-2}$  & -- & -- \\
			S4  & 5  & $2\,\mathbf{(1,2)}_{1/2} \oplus 2\,\mathbf{(1,1)}_{-1} \oplus \mathbf{(\overline{6},1)}_{2/3} \oplus \mathbf{(6,2)}_{-1/6} \oplus \mathbf{(\overline{6},1)}_{-1/3}$  & -- & -- \\
			S5  & 5  & $\mathbf{(1,2)}_{1/2} \oplus \mathbf{(1,1)}_{1} \oplus \mathbf{(1,3)}_{1} \oplus 3\,\mathbf{(1,2)}_{-3/2} \oplus 2\,\mathbf{(1,1)}_{2}$  & -- & --\\
			S6  & 5  & $ 2\,\mathbf{(1,2)}_{1/2} \oplus 3\,\mathbf{(1,1)}_{-1} \oplus \mathbf{(1,3)}_{-1} \oplus 2\,\mathbf{(1,2)}_{3/2} \oplus \mathbf{(1,1)}_{-2}$  & -- & -- \\
			S7  & 5  & $2\,\mathbf{(1,2)}_{1/2} \oplus 2\,\mathbf{(3,1)}_{-1/3} \oplus 2\,\mathbf{(1,1)}_{-1} \oplus 2\,\mathbf{(3,1)}_{2/3} \oplus 2\,\mathbf{(\overline{3},2)}_{-1/6}$ & \checkmark & -- \\
			S8  & 5  & $2\,\mathbf{(1,1)}_{-1} \oplus 2\,\mathbf{(1,3)}_{1} \oplus 2\,\mathbf{(8,1)}_{1} \oplus 2\,\mathbf{(8,2)}_{-1/2} \oplus 2\,\mathbf{(1,1)}_{-2}$& \checkmark & -- \\
			S9  & 6  & $\mathbf{(\overline{3},1)}_{1/3} \oplus \mathbf{(\overline{3},2)}_{-1/6} \oplus \mathbf{(3,3)}_{2/3} \oplus \mathbf{(3,3)}_{-1/3} \oplus \mathbf{(\overline{3},1)}_{4/3} \oplus\mathbf{(\overline{3},2)}_{-7/6}$  & $\ast$ & -- \\
			S10 & 6  & $\mathbf{(\overline{3},1)}_{1/3} \oplus \mathbf{(\overline{3},1)}_{-2/3} \oplus \mathbf{(8,1)}_{-1} \oplus \mathbf{(3,3)}_{-1/3} \oplus \mathbf{(\overline{3},1)}_{4/3} \oplus \mathbf{(8,2)}_{1/2}$  & \checkmark   & \checkmark \\
			S11 & 6  & $\mathbf{(3, 1)}_{2/3} \oplus \mathbf{(\overline{3}, 2)}_{-1/6} \oplus \mathbf{(3, 3)}_{2/3} \oplus \mathbf{(8, 1)}_{1} \oplus \mathbf{(\overline{3}, 2)}_{-7/6} \oplus \mathbf{(8,2)}_{-1/2}$  & \checkmark   & \checkmark\\
			S12 & 6   & $\mathbf{(1,2)}_{1/2} \oplus \mathbf{(6, 1)}_{-2/3} \oplus \mathbf{(\overline{6},2)}_{1/6} \oplus \mathbf{(1,2)}_{-3/2} \oplus \mathbf{(6,1)}_{1/3} \oplus \mathbf{(1,1)}_{2}$  & -- & -- \\
			S13 & 6   & $\mathbf{(\overline{6},1)}_{2/3} \oplus 2\,\mathbf{(8,1)}_{1} \oplus \mathbf{(\overline{6},1)}_{-1/3} \oplus 2\,\mathbf{(8,2)}_{-1/2} \oplus \mathbf{(6,3)}_{1/3} \oplus \mathbf{(\overline{6},1)}_{-4/3}$  & \checkmark & \checkmark \\
			S14 & 6   & $2\,\mathbf{(\overline{3},1)}_{1/3} \oplus 2\,\mathbf{(3,3)}_{2/3} \oplus \mathbf{(6,2)}_{-1/6} \oplus 2\,\mathbf{(\overline{3},2)}_{-7/6} \oplus \mathbf{(\overline{6},3)}_{-1/3} \oplus \mathbf{(6,1)}_{4/3}$ & \checkmark & \checkmark \\
			S15 & 6   & $2\,\mathbf{(\overline{3},1)}_{1/3} \oplus 2\,\mathbf{(\overline{3},1)}_{-2/3} \oplus 2\,\mathbf{(3,2)}_{1/6} \oplus \mathbf{(\overline{6},1)}_{2/3} \oplus \mathbf{(6,2)}_{-1/6} \oplus \mathbf{(\overline{6},1)}_{-1/3}$  & \checkmark & \checkmark \\
			S16 & 6  & $2\,\mathbf{(\overline{3},2)}_{-1/6} \oplus \mathbf{(\overline{6},2)}_{1/6} \oplus 2\,\mathbf{(3,3)}_{-1/3} \oplus 2\,\mathbf{(\overline{3},1)}_{4/3} \oplus \mathbf{(6,3)}_{1/3} \oplus \mathbf{(\overline{6},1)}_{-4/3}$  & \checkmark & \checkmark \\
			S17 & 6  & $\mathbf{(1,2)}_{1/2} \oplus 2\,\mathbf{(\overline{3},1)}_{1/3} \oplus 2\,\mathbf{(\overline{3},1)}_{-2/3} \oplus 2\,\mathbf{(3,2)}_{1/6} \oplus \mathbf{(1,2)}_{-3/2} \oplus \mathbf{(1,1)}_{2}$  & \checkmark & -- \\
			S18 & 6  & $\mathbf{(1,2)}_{1/2} \oplus 2\,\mathbf{(1,3)}_{1} \oplus 3\,\mathbf{(1,2)}_{-3/2} \oplus \mathbf{(8,1)}_{1} \oplus \mathbf{(\overline{8},2)}_{-1/2} \oplus \mathbf{(1,1)}_{2}$  & -- & -- \\
			S19 & 6  & $3\,\mathbf{(1,2)}_{1/2} \oplus 3\,\mathbf{(1,1)}_{-1} \oplus \mathbf{(1,3)}_{1} \oplus \mathbf{(1,2)}_{-3/2} \oplus \mathbf{(8,1)}_{1} \oplus \mathbf{(\overline{8},2)}_{-1/2}$  & $\ast$ & -- \\
			S20 & 6  & $2\,\mathbf{(1,2)}_{1/2} \oplus 2\,\mathbf{(1,1)}_{-1} \oplus 2\,\mathbf{(1,3)}_{-1} \oplus 2\,\mathbf{(1,2)}_{3/2} \oplus \mathbf{(\overline{8},1)}_{-1} \oplus \mathbf{(8,2)}_{1/2}$  & $\ast$ & -- \\
		\end{tabular}
	\end{ruledtabular}
\end{table*}

In Table~\ref{tab:Content}, we present the minimal sets of chiral multiplets, with a maximum of $n_s=6$ number of species and up to ten multiplets per set, that are anomaly free and, at the same time, lead to vectorlike states at low energies. It is worth noticing that the sets S2 and S7 correspond to one and two additional SM generations, respectively.  We shall search among the sets in Table~\ref{tab:Content} for solutions that lead to a successful gauge coupling unification at GUT and string scales.

Any self-contained unification scenario must include scalars in order to obtain the proper symmetry breaking.  In our minimal $\mathsf{SU}(5)$-inspired setup, we shall assume that the breaking of $\mathsf{SU}(5)$ into the SM group occurs at the GUT (string) scale and it is achieved through the usual $\mathsf{24}$ adjoint scalar representation. The breaking of the SM gauge group is then realized via the usual vacuum expectation value of the Higgs field in the  $\mathsf{5}$ representation.  Thus, the scalar content is $\Sigma_3\sim\mathbf{(1,3)}_0$, $\Sigma_8\sim\mathbf{(8,1)}_0$, $(X,Y)^ {\intercal}\sim\mathbf{(3,2)}_{-5/6}$, $H^{\text{\sc c}}\sim\mathbf{(3,1)}_{-1/3}$, and the Higgs doublet $H\sim\mathbf{(1,2)}_{1/2}$. Since the scalars $X$, $Y$, and $H^{\text{\sc c}}$ can dangerously mediate proton decay, in what follows we assume that their masses are of the order of the unification scale. Therefore, from the RGE viewpoint,  the only relevant scalars are  $\Sigma_3$, $\Sigma_8$, and $H$. While the mass of the Higgs doublet is required to be at the electroweak scale, the mass scales of $\Sigma_3$ and $\Sigma_8$ are allowed to vary from $M_Z$ up to the unification scale. Without any further assumption, the latter  are expected to be close to each other, i.e., $M_{\Sigma_3}\approx M_{\Sigma_8}$.

We proceed as in the previous section and make use of Eqs.~\eqref{eq:r1r2GUT}-\eqref{eq:unifscale}. We randomly vary the running weights $r_2$ to $r_{21}$ in their allowed range $[0,1]$, together with the running weights of the scalars $\Sigma_3$ and $\Sigma_8$. We then determine $r_1$ using Eq.~\eqref{eq:r1r2GUT} and calculate the unification scale $\Lambda$ through Eq.~\eqref{eq:unifscale}.

\begin{table*}
\centering
\caption{\label{tab:unif1} Chiral sets for which gauge coupling unification is achieved among the $\mathsf{SU}(5)$-inspired anomaly-free solutions given in Table~\ref{tab:Content}. The minimal scale for unification is set at $5.0\times 10^{15}$~GeV. The values of $\alpha_U^{-1}$, maximum value of the unification scale $\Lambda_\text{max}$, as well as the minimum and maximum values of the intermediate mass scales are also given.}

\begin{ruledtabular}
\begin{tabular}{ccclcclcc}
 &  &  & \multicolumn{6}{c}{Intermediate mass scales [GeV]} \\
Set & $\alpha^{-1}_{U}$ & $\Lambda_\text{max}$[GeV] &  \multicolumn{1}{c}{rep} &  \multicolumn{1}{c}{min} &  \multicolumn{1}{c}{max} &
\multicolumn{1}{c}{rep} &  \multicolumn{1}{c}{min} &  \multicolumn{1}{c}{max}\\
\hline 
S7 & $[30.5,37.6]$ & $1.0\times 10^{17}$ &  $\mathbf{(1,2)}_{1/2}$ & $M_Z$ & $7.6\times 10^{16}$ & $\mathbf{(3,1)}_{-1/3}$ & $M_Z$ & $1.3\times 10^{16}$\\
& &  & $\mathbf{(1,1)}_{-1}$ & $ 8.0\times 10^{3}$ & $7.8\times 10^{16}$ &  $\mathbf{(3,1)}_{2/3}$ &$ 2.6\times 10^{4}$ & $9.4\times 10^{16}$\\
& &  & $\mathbf{(\overline{3},2)}_{-1/6}$ & $ M_Z$ & $5.8\times10^{7}$ & & \\
\\
S8 & $[1.4,18.8]$ & $5.0\times 10^{17}$ &  $\mathbf{(1,1)}_{-1}$ & $2.2\times 10^{3}$ & $2.8\times 10^{17}$ & $\mathbf{(1,3)}_{1}$ & $ M_Z$ & $4.0\times 10^{5}$\\
& &  & $\mathbf{(8,1)}_{1}$ & $ 8.9\times 10^{14}$ & $4.7\times 10^{17}$ & $\mathbf{(8,2)}_{-1/2}$ & $ M_Z$ & $6.2\times 10^{7}$\\
& &  & $\mathbf{(1,1)}_{-2}$ & $ 9.4\times 10^{12}$ & $3.4\times 10^{17}$\\
\\
S10 & $[8.0,35.0]$ & $5.3\times 10^{17}$ &  $\mathbf{(\overline{3},1)}_{1/3}$ & $ M_Z$ & $4.2\times 10^{17}$ & $\mathbf{(\overline{3},1)}_{-2/3}$ & $ M_Z$ & $4.9\times 10^{17}$\\
& & & $\mathbf{(8,1)}_{-1}$ & $ 6.7\times 10^{3}$ & $5.0\times 10^{17}$& $\mathbf{(3,3)}_{-1/3}$ & $ M_Z$ & $1.1\times 10^{12}$\\
& & & $\mathbf{(\overline{3},1)}_{4/3}$ & $ M_Z$ & $5.0\times 10^{17}$ & $\mathbf{(8,2)}_{1/2}$ & $ M_Z$ & $4.7\times 10^{13}$\\
\\
S11 & $[4.0,34.5]$ & $5.3\times 10^{17}$ &  $\mathbf{(3, 1)}_{2/3}$ & $ M_Z$ & $4.9\times 10^{17}$ & $\mathbf{(\overline{3}, 2)}_{-1/6}$ & $ M_Z$ & $2.7\times 10^{17}$\\
& & & $\mathbf{(3, 3)}_{2/3}$ & $ M_Z$ & $9.7\times 10^{14}$ & $\mathbf{(8, 1)}_{1}$ & $ 8.4\times 10^{3}$ & $4.6\times 10^{17}$\\
& & & $\mathbf{(\overline{3}, 2)}_{-7/6}$ & $ 4.0\times 10^{5}$ & $5.1\times 10^{17}$ & $\mathbf{(8,2)}_{-1/2}$ & $M_Z$ & $4.3\times 10^{13}$\\
\\
S13 & $[1.0,35.5]$ & $5.3\times 10^{17}$ &  $\mathbf{(\overline{6},1)}_{2/3}$ & $7.4\times 10^{4}$ & $5.1\times 10^{17}$ & $\mathbf{(8,1)}_{1}$ & $ 1.9\times 10^{7}$ & $5.0\times 10^{17}$\\
& & & $\mathbf{(\overline{6},1)}_{-1/3}$ & $ 5.5\times 10^{7} $ & $ 5.0\times 10^{17}$ & $\mathbf{(8,2)}_{-1/2}$ & $ 1.4\times 10^{13} $ & $ 5.2\times 10^{17}$\\
& & & $\mathbf{(6,3)}_{1/3}$ & $ M_Z$ & $7.1\times 10^{13}$ & $\mathbf{(\overline{6},1)}_{-4/3}$ & $ M_Z$ & $5.0\times 10^{17}$\\
\\
S14 & $[1.0, 35.9]$ & $5.3\times 10^{17}$ & $\mathbf{(\overline{3},1)}_{1/3}$ & $ M_Z$ & $4.7\times 10^{17}$ & $\mathbf{(3,3)}_{2/3}$ & $ 6.4\times 10^{7}$ & $4.9\times 10^{17}$\\
& & & $\mathbf{(6,2)}_{-1/6}$ & $M_Z$ & $ 2.8\times 10^{16}$ & $\mathbf{(\overline{3},2)}_{-7/6}$ & $ 2.9\times 10^{3}$ & $5.0\times 10^{17}$\\
& & & $\mathbf{(\overline{6},3)}_{-1/3}$ & $2.5\times 10^{3}$ & $4.4\times 10^{17}$ & $\mathbf{(6,1)}_{4/3}$ & $ M_Z$ & $4.9\times 10^{17}$\\
\\
S15 & $[32.3,37.3]$ & $5.3\times 10^{17}$ &  $\mathbf{(\overline{3},1)}_{1/3}$ & $ 9.0\times 10^{6}$ & $4.3\times 10^{17}$ & $\mathbf{(\overline{3},1)}_{-2/3}$ & $ 1.5\times 10^{11}$ & $4.3\times 10^{17}$\\
& &  & $\mathbf{(3,2)}_{1/6}$ & $ M_Z$ & $9.8\times 10^{5}$ & $\mathbf{(\overline{6}, 1)}_{2/3}$ & $ 2.1\times 10^{13}$ & $ 5.0\times 10^{17}$\\
& &  & $\mathbf{(6,2)}_{-1/6}$ & $ 8.1\times 10^{10}$ & $ 8.5\times 10^{16}$ & $\mathbf{(\overline{6}, 1)}_{-1/3}$ & $ 5.4\times 10^{12}$ & $4.8\times 10^{17}$\\
\\
S16 & $[1.0, 37.1]$ & $5.3\times 10^{17}$ & $\mathbf{(\overline{3},2)}_{-1/6}$ & $ M_Z$ & $ 5.1\times 10^{17}$ & $\mathbf{(\overline{6},2)}_{1/6}$ & $ M_Z$ & $6.0\times 10^{15}$\\
& & & $\mathbf{(3,3)}_{-1/3}$ & $ 9.5\times 10^{7}$ & $5.3\times 10^{17}$ & $\mathbf{(\overline{3},1)}_{4/3}$ & $ M_Z$ & $5.0\times 10^{17}$\\
& & & $\mathbf{(6,3)}_{1/3}$ & $ 880$ & $4.9\times 10^{17}$ & $\mathbf{(\overline{6},1)}_{-4/3}$ & $ M_Z$ & $5.0\times 10^{17}$\\
\\
S17 & $[31.9,37.1]$ & $6.0\times 10^{16}$ &  $\mathbf{(1,2)}_{1/2}$ & $M_Z$ & $2.0\times 10^{16}$ & $\mathbf{(\overline{3},1)}_{1/3}$  & $M_Z$ & $8.3\times 10^{15}$\\
& & & $\mathbf{(\overline{3},1)}_{-2/3}$ & $ 2.9\times 10^{6}$ & $3.3\times 10^{16}$ & $\mathbf{(3,2)}_{1/6}$  & $ M_Z$ & $2.0\times 10^{7}$\\
& & & $\mathbf{(1,2)}_{-3/2}$ & $ 6.5\times 10^{10}$ & $ 5.0\times 10^{16}$ & $\mathbf{(1,1)}_{2}$ & $6.7\times 10^{10}$ & $5.9\times 10^{16}$\\
\end{tabular}
\end{ruledtabular}
\end{table*}

\begin{figure}[b]
	\centering
	\includegraphics[width=0.95\linewidth]{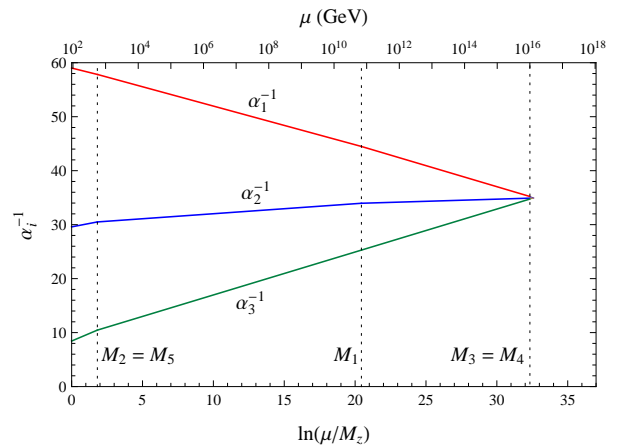}
	\caption{\label{fig:S7GUT} Running of the gauge couplings at one-loop level for the set S7 given in Table~\ref{tab:unif1}, which corresponds to the addition of two complete SM generations. A successful unification of the couplings is achieved at the scale $\Lambda = 1.3\times10^{16}$~GeV. We assume $M_{\Sigma_3}= M_{\Sigma_8}=\Lambda$.}
\end{figure}

\begin{table*}
	\centering
	\caption{\label{tab:unif2} As in Table~\ref{tab:unif1}, for the sets of solutions that lead to gauge and gravitational coupling unification at the string scale.}
	
	\begin{ruledtabular}
		\begin{tabular}{cclcclcc}
			&  & \multicolumn{6}{c}{Intermediate mass scales [GeV]} \\
			Set & $\Lambda\,[10^{17}\,\text{GeV}]$ &  \multicolumn{1}{c}{rep} &  \multicolumn{1}{c}{min} &  \multicolumn{1}{c}{max} &
			\multicolumn{1}{c}{rep} &  \multicolumn{1}{c}{min} &  \multicolumn{1}{c}{max}\\
			\hline
			S10 & $[3.5,5.3]$ &  $\mathbf{(\overline{3},1)}_{1/3}$ & $ M_Z$ & $5.0\times 10^{17}$ & $\mathbf{(\overline{3},1)}_{-2/3}$ & $ M_Z$ & $4.6\times 10^{17}$\\
			& & $\mathbf{(8,1)}_{-1}$ & $ 2.9\times 10^{7}$ & $4.5\times 10^{17}$& $\mathbf{(3,3)}_{-1/3}$ & $ 3.5\times 10^3$ & $2.4\times 10^{12}$\\
			& & $\mathbf{(\overline{3},1)}_{4/3}$ & $ 9.8\times 10^{7}$ & $5.2\times 10^{17}$ & $\mathbf{(8,2)}_{1/2}$ & $ M_Z$ & $1.1\times 10^{8}$\\
			\\
			S11 & $[3.6,5.3]$ &  $\mathbf{(3, 1)}_{2/3}$ & $M_Z$ & $4.3\times 10^{17}$ & $\mathbf{(\overline{3}, 2)}_{-1/6}$ & $ M_Z$ & $8.8\times 10^{16}$\\
			& & $\mathbf{(3, 3)}_{2/3}$ & $9.6\times 10^{7}$ & $3.0\times 10^{14}$ & $\mathbf{(8, 1)}_{1}$ & $1.3\times 10^{11}$ & $4.4\times 10^{17}$\\
			& & $\mathbf{(\overline{3}, 2)}_{-7/6}$ & $5.7\times 10^{12}$ & $5.0\times 10^{17}$ & $\mathbf{(8,2)}_{-1/2}$ & $M_Z$ & $1.1\times 10^{8}$\\
			\\
			S13 & $[3.3,5.3]$ &  $\mathbf{(\overline{6},1)}_{2/3}$ & $ 7.9\times 10^{3}$ & $5.2\times 10^{17}$ & $\mathbf{(8,1)}_{1}$ & $7.7\times 10^{9}$ & $5.2\times 10^{17}$\\
			& & $\mathbf{(\overline{6},1)}_{-1/3}$ & $3.1\times 10^{6}$ & $5.2\times 10^{17}$ & $\mathbf{(8,2)}_{-1/2}$ & $8.3\times 10^{12}$ & $5.2\times 10^{17}$\\
			& & $\mathbf{(6,3)}_{1/3}$ & $4.8\times 10^{5}$ & $1.3\times 10^{14}$ & $\mathbf{(\overline{6},1)}_{-4/3}$ & $3.3\times 10^{6}$ & $5.0\times 10^{17}$\\
			\\
			S14 & $[3.2,5.3]$ & $\mathbf{(\overline{3},1)}_{1/3}$ & $ M_Z$ & $5.0\times 10^{17}$ & $\mathbf{(3,3)}_{2/3}$ & $7.6\times 10^{10}$ & $5.2\times 10^{17}$\\
			& & $\mathbf{(6,2)}_{-1/6}$ & $M_Z$ & $3.3\times 10^{15}$ & $\mathbf{(\overline{3},2)}_{-7/6}$ & $5.2\times 10^{9}$ & $5.1\times 10^{17}$\\
			& & $\mathbf{(\overline{6},3)}_{-1/3}$ & $2.3\times 10^{7}$ & $4.3\times 10^{17}$ & $\mathbf{(6,1)}_{4/3}$ & $5.8\times 10^{5}$ & $5.2\times 10^{17}$\\
			\\
			S15 & $[3.2,3.3]$ &  $\mathbf{(\overline{3},1)}_{1/3}$ & $1.3\times 10^{7}$ & $3.3\times 10^{17}$ & $\mathbf{(\overline{3},1)}_{-2/3}$ & $1.7\times 10^{11}$ & $3.2\times 10^{17}$\\
			& & $\mathbf{(3,2)}_{1/6}$ & $M_Z$ & $8.4\times 10^{5}$ & $\mathbf{(\overline{6}, 1)}_{2/3}$ & $1.0\times 10^{14}$ & $3.2\times 10^{17}$\\
			& & $\mathbf{(6,2)}_{-1/6}$ & $1.8\times 10^{11}$ & $5.2\times 10^{16}$ & $\mathbf{(\overline{6}, 1)}_{-1/3}$ & $7.8\times 10^{12}$ & $3.2\times 10^{17}$\\
			\\
			S16 & $[3.2,5.3]$ & $\mathbf{(\overline{3},2)}_{-1/6}$ & $M_Z$ & $5.2\times 10^{17}$ & $\mathbf{(\overline{6},2)}_{1/6}$ & $ M_Z$ & $3.6\times 10^{12}$\\
			& & $\mathbf{(3,3)}_{-1/3}$ & $ 1.7\times 10^{11}$ & $5.3\times 10^{17}$ & $\mathbf{(\overline{3},1)}_{4/3}$ & $8.5\times 10^{4}$ & $5.2\times 10^{17}$\\
			& & $\mathbf{(6,3)}_{1/3}$ & $1.8\times 10^{8}$ & $5.2\times 10^{17}$ & $\mathbf{(\overline{6},1)}_{-4/3}$ & $1.1\times 10^{5}$ & $5.2\times 10^{17}$\\
		\end{tabular}
	\end{ruledtabular}
\end{table*}

In Table~\ref{tab:unif1}, we list the nine sets of solutions that lead to unification of the gauge couplings for $\Lambda \geq 5\times10^{15}$~GeV. As mentioned before, this lower bound is invoked so that we are safely consistent with proton decay bounds.  In fact, we have also found unification of the gauge couplings for the sets S9, S19, and S20, but the unification scale $\Lambda$ turns out to be very constrained for those sets ($\Lambda\lesssim10^ {16}$ GeV). The values of $\alpha_U^{-1}$, the maximum value of the unification scale $\Lambda_\text{max}$, as well as the minimum and maximum values of the intermediate mass scales obtained for each set are also given in Table~\ref{tab:unif1}. In all cases, the mass scale $M_{\Sigma_3} \simeq M_{\Sigma_8}$ can take any value from  $M_Z$ up to the unification scale.

As can be seen from Table~\ref{tab:unif1}, it is possible to achieve unification with some of the intermediate scales taking values as low as the electroweak scale. As an illustration, we present in Fig.~\ref{fig:S7GUT} the running of the gauge couplings at one-loop level for the set S7, which corresponds to the addition of two complete SM generations. In this example, the unification of the gauge couplings occurs at the scale $\Lambda = 1.3\times10^{16}$ GeV with $\alpha^{-1}_{U} \simeq 35$. The intermediate mass scales are $M_2 = M_5 = 560$~GeV, $M_1 = 6.9\times10^{10}$~GeV, and $M_3 = M_4 = 1.0\times10^{16}$~GeV. Note that the mass labeling $M_i$, with $i=1, \ldots, 21$, follows the multiplet notation given in Table~\ref{tab:ParticleBis}. We assume $M_{\Sigma_3}= M_{\Sigma_8}=\Lambda$.

Let us now consider the string scenario. In order to determine if the gauge and gravitational couplings unify in this case, we compute, for each set of solutions, the running weights of the first two particles, $r_1$ and $r_2$, using Eq.~\eqref{eq:ristring}, allowing the running weights of the remaining particles to randomly vary in the range $[0,1]$. The unification scale $\Lambda$ is determined using Eq.~\eqref{eq:lamber}. We have found only six sets compatible with string-scale unification, which are presented in Table~\ref{tab:unif2}. One can see that, in order to have unification, the presence of sextets and/or octets of $\mathsf{SU}(3)$ is required in all cases.

\begin{table}[b]
\centering
\caption{\label{tab:particleSU5} The cubic index $A_5$  for irreducible representations of $\mathsf{SU}(5)$ with $d_5(R)\leq50$.}

\begin{ruledtabular}
\begin{tabular}{ccccccccccccccccc}
$\mathsf{SU}(5)$-irrep  & & $\mathsf{5}$ & & $\mathsf{10}$ & & $\mathsf{15}$ & & $\mathsf{24}$ & &  $\mathsf{35}$ & & $\mathsf{40}$ & & $\mathsf{45}$ & & $\mathsf{50}$\\
\hline
$A_5$ & & 1 & & 1 & & 9 & & 0 & & -44 & & -16 & & -6 & & -15 \\
\end{tabular}
\end{ruledtabular}
\end{table}

We conclude this section by commenting on the possibility of constructing anomaly-free solutions with complete $\mathsf{SU(5)}$ representations. Looking at Table~\ref{tab:Content}, one can easily verify that none of the sets corresponds to a complete representation of this gauge group. In order to obtain anomaly cancellation within $\mathsf{SU(5)}$, we must consider the $\mathsf{SU(5)}$ anomaly condition
\begin{equation}
\label{eq:su5anomaly}
\left[\mathsf{SU}(5)\,\mbox{-}\,\mathsf{SU}(5)\,\mbox{-}\, \mathsf{SU}(5)\right]:\quad \sum_R A_5(R)\,=\,0\,,
\end{equation}
where the anomaly cubic index $A_5(R)$ is given in Table~\ref{tab:particleSU5}. For our choice of $\mathsf{SU}(5)$ representations, with dimensions less than or equal to 50, the above equation implies the following relation among the number of $\mathsf{SU}(5)$ multiplets:
\begin{equation}
\label{eq:su5anomalyrel}
\mathsf{n_{5}} \,+\, \mathsf{n_{10}}  \,+\,  9 \mathsf{n_{15}}  \,-\, 44  \mathsf{n_{35}}  \,-\, 16 \mathsf{n_{40}} \,-\, 6  \mathsf{n_{45}} \,-\, 15 \mathsf{n_{50} }\,=\,0\,.
\end{equation}
Moreover, imposing the low-energy vectorlike conditions for the particle content of the above multiplets, we obtain the relations
\begin{equation}
\label{eq:su5vectorlike}
\begin{aligned}
&\mathsf{n_{5}} \,+\, \mathsf{n_{10}} \,-\,  \mathsf{n_{40}} \,=\,0\,,\\
&\mathsf{n_{15}}  \,-\, \mathsf{n_{40}}  \,-\, \mathsf{n_{50}}\,=\,0\,,\\
&\mathsf{n_{15}} \,+\, \mathsf{n_{45}} \,=\,0\,,\\
&\mathsf{n_{35}} \,=\,0\,.
\end{aligned}
\end{equation}
One can easily verify that once Eqs.~\eqref{eq:su5vectorlike} are imposed, Eq.~\eqref{eq:su5anomaly} is automatically satisfied.

Besides the well-known SM solution, formed by the combination $\overline{\mathsf{5}} \oplus \mathsf{10}$, we have found  the following minimal anomaly-free sets of complete representations of $\mathsf{SU}(5)$,
\begin{equation}
\label{eq:smNewWay}
\begin{aligned}
\overline{\mathsf{5}}\,& \oplus \, \overline{\mathsf{40}} \, \oplus \, \mathsf{50}\,,\\
\mathsf{10} \, & \oplus \, \mathsf{40} \, \oplus \, \overline{\mathsf{50}}\,,\\
\mathsf{15} \, &\oplus \, \overline{\mathsf{45}} \, \oplus \, \mathsf{50}\,,\\
\overline{\mathsf{5}} \,& \oplus \, \overline{\mathsf{15}} \, \oplus \, \overline{\mathsf{40}} \, \oplus \, \mathsf{45}\,,\\
\mathsf{10} \,& \oplus \, \mathsf{15} \, \oplus \, \mathsf{40} \, \oplus \, \overline{\mathsf{45}}\,,
\end{aligned}
\end{equation}
which contain at most three or four complete $\mathsf{SU}(5)$ representations. It is interesting to note that the last anomaly-free chiral set contains, in a nontrivial way, one generation of the SM fermions (labeled 1-5 in Table~\ref{tab:ParticleBis}).

Concerning gauge coupling unification, it is clear that, unless one allows for mass splittings inside each $\mathsf{SU}(5)$ multiplet, none of the solutions in Eq.~\eqref{eq:smNewWay} would lead to unification. Indeed, the contribution of the new particles [belonging to a complete $\mathsf{SU}(5)$ multiplet] to the beta coefficients at one loop in Eqs.~\eqref{eq:iasolution}-\eqref{eq:ri} is the same for each SM subgroup. Therefore, they do not affect the unification condition of Eq.~\eqref{eq:unifgut}. On the other hand, if one allows for different intermediate scales inside each $\mathsf{SU}(5)$ multiplet, one can then obtain sets of solutions that can account for unification but they are not as minimal as those given in Table~\ref{tab:unif1}.

\section{Conclusions}
\label{sec:conclusions}

In this work we have searched for minimal chiral sets of fermions beyond the SM that are anomaly free and, simultaneously, vectorlike particles with respect to color $\mathsf{SU}(3)$ and electromagnetic $\mathsf{U}(1)$. We have studied whether the addition of such particles allows for the unification of gauge couplings at a high energy scale. The possibility to have unification at the string scale has also been considered. We have looked for minimal solutions with chiral fermion sets of arbitrary quantum numbers, which do no fit in standard GUT groups, as well as for those that belong to $\mathsf{SU}(5)$ representations with dimensions less than or equal to~$50$. In both cases, we have verified that some of the anomaly-free sets can unify at a unification scale above $5 \times10^{15}$ GeV and could also lead to the unification of gauge and gravitational couplings at the string scale around $3 \times 10^{17}$~GeV. Our results are summarized in Figs.~\ref{fig:P1a3}-\ref{fig:P4}, and Tables~\ref{tab:paradigmaString}, \ref{tab:unif1}, and \ref{tab:unif2}. 

In our framework, we have only considered one-loop corrections to the running of the gauge couplings, allowing the extra fermions to decouple from the theory at arbitrary intermediate scales. A more comprehensive analysis would require us to include higher loop RGE corrections. Our results show that adding only a minimal chiral content to SM and requiring gauge unification enforces that some of the new particles should decouple from the theory at intermediate scales much larger than the electroweak scale. Since electroweak precision data severely constrain additional scalars charged under the SM group, giving large masses to the new fermions through such scalars seems unrealistic in the present context. Therefore, the problem of generating the required intermediate fermion masses through an alternative mechanism deserves further study. An interesting possibility is to implement a dynamical mechanism by extending the theory with mirror fermions, analogous to that developed in Ref.~\cite{Lizzi:1997sg}.

Finally, it is worth emphasizing that our search for gauge coupling unification has been focused on nonsupersymmetric scenarios. Supersymmetry commonly arises in the context of string theories; yet it may happen that it is broken at a very high energy scale. In fact, different paths to string-scale unification can be envisaged~\cite{Dienes:1996du}. Supersymmetry may also be required to maintain the stability of the relevant mass scales, namely, the string and/or gauge unification scale, the intermediate scales for the masses of the new fermions, and the electroweak scale. For instance, invoking low-energy supersymmetry helps in solving the hierarchy problem associated to the presence of quadratic divergences in the Higgs mass. So far, supersymmetry has not been observed at the energy scales that are accessible to present collider experiments. In view of this, nonsupersymmetric extensions of the SM remain plausible alternatives that are worth being investigated.   

\begin{acknowledgments}
D.E.C. is grateful to P.B. Pal for useful discussions. The work of L.M.C. was financially supported by Centro de F\'{\i}sica
  Te\'{o}rica de Part\'{\i}culas (CFTP) through the research Grant No. BL/67/2014. D.E.C. thanks the CERN Theory Division for hospitality and financial support. The work of D.E.C.~was supported by Associa\c{c}\~{a}o do Instituto Superior T\'{e}cnico para a Investiga\c{c}\~{a}o e Desenvolvimento (IST-ID) and Funda\c{c}\~{a}o para a Ci\^{e}ncia e a Tecnologia (FCT) through the Grants No. PEst-OE/FIS/UI0777/2013, No. CERN/FP/123580/2011, and No. PTDC/FIS-NUC/0548/2012. C.S. thanks CFTP for the hospitality. The work of R.G.F. was partially supported by FCT, under the Grants No. PEst-OE/FIS/UI0777/2013 and No. CERN/FP/123580/2011.
\end{acknowledgments}

\bibliography{refs}

\end{document}